\documentclass[aps,prd,nofootinbib,twocolumn,superscriptaddress,preprintnumbers,balancelastpage,longbibliography,nobibnotes]{revtex4-2}

\usepackage{graphicx,epsfig,psfrag,bm,amssymb,adjustbox}
\usepackage{dcolumn}
\usepackage{bm}
\usepackage[dvipsnames]{xcolor}
\usepackage{braket}
\usepackage{mathrsfs,amsfonts,hepunits,color}
\usepackage{hyperref}
\usepackage{comment}
\usepackage{physics}
\usepackage{subcaption}
\usepackage{siunitx}

\begin{document}

\title{Quantum computing hardware for HEP algorithms and sensing}

\newcommand\snowmass{\begin{center}\rule[-0.2in]{\hsize}{0.01in}\\\rule{\hsize}{0.01in}\\
\vskip 0.1in Submitted to the  Proceedings of the US Community Study\\
on the Future of Particle Physics (Snowmass 2021)\\
\rule{\hsize}{0.01in}\\\rule[+0.2in]{\hsize}{0.01in} \end{center}}

\author{M. Sohaib Alam} 
\affiliation{Quantum Artificial Intelligence Laboratory (QuAIL), NASA Ames Research Center, Moffett Field, CA, 94035, USA}
\affiliation{USRA Research Institute for Advanced Computer Science (RIACS), Mountain View, CA, 94043, USA}

\author{Sergey Belomestnykh}
\affiliation{Superconducting Quantum Materials and Systems Center (SQMS), Fermi National Accelerator Laboratory, Batavia, IL 60510, USA}

\author{Nicholas Bornman}
\affiliation{Superconducting Quantum Materials and Systems Center (SQMS), Fermi National Accelerator Laboratory, Batavia, IL 60510, USA}

\author{Gustavo Cancelo}
\affiliation{Scientific Computing Division, Fermi National Accelerator Laboratory, Batavia, IL, 60510, United States of America}

\author{Yu-Chiu Chao}
\affiliation{Superconducting Quantum Materials and Systems Center (SQMS), Fermi National Accelerator Laboratory, Batavia, IL 60510, USA}

\author{Mattia Checchin}
\affiliation{Superconducting Quantum Materials and Systems Center (SQMS), Fermi National Accelerator Laboratory, Batavia, IL 60510, USA}

\author{Vinh San Dinh}
\affiliation{Superconducting Quantum Materials and Systems Center (SQMS), Fermi National Accelerator Laboratory, Batavia, IL 60510, USA}
\affiliation{Northwestern-Fermilab Center for Applied Physics and Superconducting Technologies,
Northwestern University, Evanston, Illinois 60208, USA}
\affiliation{Graduate Program in Applied Physics, Northwestern University, Evanston, Illinois 60208, USA}

\author{Anna Grassellino}
\affiliation{Superconducting Quantum Materials and Systems Center (SQMS), Fermi National Accelerator Laboratory, Batavia, IL 60510, USA}

\author{Erik J. Gustafson}
\affiliation{Superconducting Quantum Materials and Systems Center (SQMS), Fermi National Accelerator Laboratory, Batavia, IL 60510, USA}
\affiliation{Theory Division, Fermi National Accelerator Laboratory, Batavia,  IL 60510, USA}

\author{Roni Harnik}
\affiliation{Superconducting Quantum Materials and Systems Center (SQMS), Fermi National Accelerator Laboratory, Batavia, IL 60510, USA}
\affiliation{Theory Division, Fermi National Accelerator Laboratory, Batavia,  IL 60510, USA}

\author{Corey Rae Harrington McRae}
\affiliation{Department of Physics and Department of Electrical, Computer and Energy Engineering, University of Colorado Boulder, Colorado 80305, USA}
\affiliation{National Institute of Standards and Technology Boulder, Colorado 80305, USA}

\author{Ziwen Huang}
\affiliation{Superconducting Quantum Materials and Systems Center (SQMS), Fermi National Accelerator Laboratory, Batavia, IL 60510, USA}

\author{Keshav Kapoor}
\affiliation{Superconducting Quantum Materials and Systems Center (SQMS), Fermi National Accelerator Laboratory, Batavia, IL 60510, USA}

\author{Taeyoon Kim}
\affiliation{Superconducting Quantum Materials and Systems Center (SQMS), Fermi National Accelerator Laboratory, Batavia, IL 60510, USA}
\affiliation{Northwestern-Fermilab Center for Applied Physics and Superconducting Technologies,
Northwestern University, Evanston, Illinois 60208, USA}
\affiliation{Department of Physics and Astronomy, Northwestern University, Evanston, Illinois 60208, USA}

\author{James B. Kowalkowski}
\affiliation{Superconducting Quantum Materials and Systems Center (SQMS), Fermi National Accelerator Laboratory, Batavia, IL 60510, USA}

\author{Matthew J. Kramer}
\affiliation{Ames Laboratory, U.S. Department of Energy, Ames, IA 50011, United States}

\author{Yulia Krasnikova}
\affiliation{Superconducting Quantum Materials and Systems Center (SQMS), Fermi National Accelerator Laboratory, Batavia, IL 60510, USA}

\author{Prem Kumar}
\affiliation{Superconducting Quantum Materials and Systems Center (SQMS), Fermi National Accelerator Laboratory, Batavia, IL 60510, USA}
\affiliation{Center for Photonic Communication and Computing, ECE Department, Northwestern University, Evanston, IL 60208, USA}

\author{Doga Murat Kurkcuoglu}
\affiliation{Superconducting Quantum Materials and Systems Center (SQMS), Fermi National Accelerator Laboratory, Batavia, IL 60510, USA}

\author{Henry~Lamm}
\affiliation{Superconducting Quantum Materials and Systems Center (SQMS), Fermi National Accelerator Laboratory, Batavia, IL 60510, USA}
\affiliation{Theory Division, Fermi National Accelerator Laboratory, Batavia,  IL 60510, USA}

\author{Adam L. Lyon}
\affiliation{Superconducting Quantum Materials and Systems Center (SQMS), Fermi National Accelerator Laboratory, Batavia, IL 60510, USA}

\author{Despina Milathianaki}
\affiliation{Rigetti Computing, Inc., Berkeley, CA 94710, USA}

\author{Akshay Murthy}
\affiliation{Superconducting Quantum Materials and Systems Center (SQMS), Fermi National Accelerator Laboratory, Batavia, IL 60510, USA}

\author{Josh Mutus}
\affiliation{Rigetti Computing, Inc., Berkeley, CA 94710, USA}

\author{Ivan Nekrashevich}
\affiliation{Superconducting Quantum Materials and Systems Center (SQMS), Fermi National Accelerator Laboratory, Batavia, IL 60510, USA}

\author{JinSu Oh}
\affiliation{Ames Laboratory, U.S. Department of Energy, Ames, IA 50011, United States}

\author{A. Bar\i\c{s} \"Ozg\"uler}
\affiliation{Superconducting Quantum Materials and Systems Center (SQMS), Fermi National Accelerator Laboratory, Batavia, IL 60510, USA}
\author{Gabriel Nathan Perdue}
\affiliation{Superconducting Quantum Materials and Systems Center (SQMS), Fermi National Accelerator Laboratory, Batavia, IL 60510, USA}

\author{Matthew Reagor}
\affiliation{Rigetti Computing, Inc., Berkeley, CA 94710, USA}

\author{Alexander Romanenko}
\affiliation{Superconducting Quantum Materials and Systems Center (SQMS), Fermi National Accelerator Laboratory, Batavia, IL 60510, USA}

\author{James A. Sauls}
\affiliation{Superconducting Quantum Materials and Systems Center (SQMS), Fermi National Accelerator Laboratory, Batavia, IL 60510, USA}
\affiliation{Northwestern-Fermilab Center for Applied Physics and Superconducting Technologies, Northwestern University, Evanston, Illinois 60208, USA}
\affiliation{Department of Physics and Astronomy, Northwestern University, Evanston, Illinois 60208, USA}

\author{Leandro Stefanazzi}
\affiliation{Scientific Computing Division, Fermi National Accelerator Laboratory, Batavia, IL, 60510, United States of America}

\author{Norm M. Tubman}
\affiliation{NASA Ames Research Center, Moffett Field, CA 94035, USA}

\author{Davide Venturelli} 
\affiliation{Quantum Artificial Intelligence Laboratory (QuAIL), NASA Ames Research Center, Moffett Field, CA, 94035, USA}
\affiliation{USRA Research Institute for Advanced Computer Science (RIACS), Mountain View, CA, 94043, USA}

\author{Changqing Wang}
\affiliation{Superconducting Quantum Materials and Systems Center (SQMS), Fermi National Accelerator Laboratory, Batavia, IL 60510, USA}

\author{Xinyuan You}
\affiliation{Superconducting Quantum Materials and Systems Center (SQMS), Fermi National Accelerator Laboratory, Batavia, IL 60510, USA}

\author{David M. T. van Zanten}
\email[email: ]{dvanzant@fnal.gov}
\affiliation{Superconducting Quantum Materials and Systems Center (SQMS), Fermi National Accelerator Laboratory, Batavia, IL 60510, USA}

\author{Lin Zhou}
\affiliation{Ames Laboratory, U.S. Department of Energy, Ames, IA 50011, United States}

\author{Shaojiang Zhu}
\affiliation{Superconducting Quantum Materials and Systems Center (SQMS), Fermi National Accelerator Laboratory, Batavia, IL 60510, USA}

\author{Silvia Zorzetti}
\email[email: ]{zorzetti@fnal.gov}
\affiliation{Superconducting Quantum Materials and Systems Center (SQMS), Fermi National Accelerator Laboratory, Batavia, IL 60510, USA}

\date{\today}

\begin{abstract} 
Quantum information science harnesses the principles of quantum mechanics to realize computational algorithms with complexities vastly intractable by current computer platforms. Typical applications range from quantum chemistry to optimization problems and also include simulations for high energy physics. The recent maturing of quantum hardware has triggered preliminary explorations by several institutions (including Fermilab) of quantum hardware capable of demonstrating quantum advantage in multiple domains, from quantum computing to communications, to sensing. The Superconducting Quantum Materials and Systems (SQMS) Center, led by Fermilab, is dedicated to providing breakthroughs in quantum computing and sensing, mediating quantum engineering and HEP based material science. The main goal of the Center is to deploy quantum systems with superior performance tailored to the algorithms used in high energy physics. In this Snowmass paper, we discuss the two most promising superconducting quantum architectures for HEP algorithms, i.e. three-level systems (qutrits) supported by transmon devices coupled to planar devices and multi-level systems (qudits with arbitrary N energy levels) supported by superconducting 3D cavities. For each architecture, we demonstrate exemplary HEP algorithms and identify the current challenges, ongoing work and future opportunities. Furthermore, we discuss the prospects and complexities of interconnecting the different architectures and individual computational nodes. Finally, we review several different strategies of error protection and correction and discuss their potential to improve the performance of the two architectures. This whitepaper seeks to reach out to the HEP community and drive progress in both HEP research and QIS hardware.


\vspace{-4.8mm}

\begin{center}
\snowmass

\end{center}

\end{abstract}

\preprint{FERMILAB-PUB-22-260-SQMS}

\maketitle

\makeatletter
\def\l@subsubsection#1#2{}
\makeatother

\tableofcontents

\section{Introduction} 

Theoretical and experimental investigation in high energy physics (HEP) has long been the prime driver for pushing the frontiers of computation paradigms and raw computing power. 
Following the Feynman's seminal paper on the quantum simulator in 1982, quantum computers provide a powerful alternative to the classical machines.
The need for algorithmic and numerical implementation of lattice field theory, and in particular lattice quantum chromodynamics, has resulted in revolutionary advances in classical computing hardware and algorithms. Despite these successes, several problems of profound interest to the HEP community are intractable upon classical computers due to the so-called \emph{sign problems}.  These obstacles can be elegantly avoided via quantum simulators, strongly motivating the HEP community's involvement in developing quantum hardware and the necessary algorithms \cite{Davoudi:2022cah}. 
Technology and expertise developed by the HEP community provide exceptional theoretical and experimental resources to advance quantum information science (QIS). The main goal of the Superconducting Quantum Materials and Systems (SQMS) center, led by Fermilab, is to mitigate decoherence mechanisms of current quantum devices and deploy superior quantum systems to advance quantum algorithms and sensing. The main advantage of the cavity-based systems explored at SQMS for the HEP simulations come from encoding the quantum state in the low energy modes of the transmon, or bosonic modes of the electric field inside the 3D resonator, which can be utilized as `qudits' with $N$ states instead of qubits with 2 states. This implementation has fewer restrictions on the number of gates that needs to be applied compared to the qubit-based devices. A parallel effort in quantum sensing is launched for dark matter and axion searches, taking advantage of our integrated environment of quantum devices, materials and superconducting radio-frequency (SRF) infrastructure \cite{Berlin:2022hfx}.

\subsection{From the theory of quantum harmonic oscillators to the engineering quantum devices}

The quantum harmonic oscillator is one of the first topics discussed in most introduction courses on quantum mechanics. The system Hamiltonian is relatively easily obtained starting from the position and momentum operators and produces a spectrum of eigenstates, the Hilbert space, of equidistant energy (spaced by the natural frequency $f_C$), known as Fock states. For an electromagnetic resonator, the solution is identical, but the position and momentum operators are replaced by the electric and magnetic fields.  Electromagnetic resonators can be classified in two types: planar circuits (2D) or bulk cavities (3D).
The effective dimension of the Hilbert space (bounded by energy dissipation) could potentially provide an enormous computational space, assuming the ability to create and manipulate any arbitrary superposition of Fock states. The homogeneity of the level spacing, resulting from the linearity of the system, however, makes controlling the photon occupation probability of each individual Fock state impossible. This is because the application of any resonant drive tone of frequency $f_D = f_C$ will lead to transitions between all levels resulting in a coherent state (i.e. a superposition of Fock states with a Poisson probability distribution). 
Introducing some non-linearity into the system changes the response strikingly. In this case the energy levels become unevenly spaced and consequently the drive tone is only resonant with a single pair of levels resulting in oscillations between just those levels, e.g. between the ground and first excited (Fock) state. This level selectivity allows for the preparation of any arbitrary superposition of Fock states and ultimately quantum information processing.  

Any quantum state, however, will in time be affected adversely by interactions with the environment. The most evident result of this is the energy relaxation, which takes place via a photon exchange with a bath of atomic two-level systems (TLS). The energy decay rate of the oscillator (among others) is determined by 1) the dielectric losses related to the electric field, and 2) the conduction losses related to magnetic fields. At the low excitation power typically used in quantum experiments (of the order of a single photon) the contribution of the dielectric losses dominates the relaxation rate. With increasing field strength, however, the relaxation rate decreases and saturates to a value defined by residual losses, which in the case of cavity oscillators is determined by the conductive losses. 
In addition to the energy relaxation, interactions with the environment may also lead to the dephasing of the quantum state. 
Therefore, quantum devices are typically engineered to have an exponentially small dispersion with respect to the most noisy system parameters, to avoid stochastic noise affecting the transition frequency 
causing dephasing of the quantum state.

Research has shown that with a large Hilbert space, encoding quantum algorithms in qutrits and qudits can often outperform their qubit counterparts in terms of complexity \cite{gokhale2019, nikolaeva2021efficient,gedik2015computational, wang2020qudits,lanyon2009simplifying,2017PhRvA..96a2306B}. This improvement is significantly driven by reduction in entangling gate depths and larger memory, and offers an attractive perspective to improve the performances of algorithms for HEP simulations. 

In quantum sensing, superconducting cavities with long-coherence can also be used as powerful quantum sensors to probe fundamental physics, overcoming the limitations of current sensing schemes, which are limited by the ability to store and detect single microwave photons.

\subsection{Organization of this whitepaper}
In Sections \ref{3DQPU} and \ref{2DQPU}, we review 3D and 2D quantum processing units (QPUs). For each system, we discuss encoding schemes based on three-level systems (qutrits) or multi-level systems (qudits), with particular emphasis on the algorithm implementations for HEP. We also evaluate materials studies for the 3D and 2D quantum systems, focusing on losses and mitigation mechanisms in bulk cavities, as well as in planar circuits. Inter-connecting quantum devices is a major challenge that needs to be addressed to deploy scalable quantum computers with both high efficiency and fidelity, with applications in quantum communications and sensing. The implementation of distributed quantum networks leads to the need of manipulating individual quantum nodes, as well as realizing hybrid devices for quantum transduction. This is discussed in Sec. \ref{Interconnectivity}, along with different tunable schemes. We also discuss room temperature hardware for quantum devices in Sec. \ref{room_hardware}, highlighting specifications and requirements to address a large Hilbert space and multi-mode quantum systems. Quantum error protection and correction is treated in Sec. \ref{error_correction}. To perform useful quantum computation with high-fidelity, the implementation of both hardware-level and active error correction strategies needs to be evaluated.  In Section \ref{HEPcloud} we discuss how to make quantum computing hardware available to the broader HEP community through a dedicated HEPCloud platform that provides standardized methods and computational resources based on the SQMS systems. Finally, in Sec. \ref{Conclusions}, we outline some of the R\&D directions in synergy between the QIS and HEP programs.

\section{3D QPU}\label{3DQPU}
\begin{figure}[!ht]
    \centering
     \includegraphics[width=0.3\textwidth]{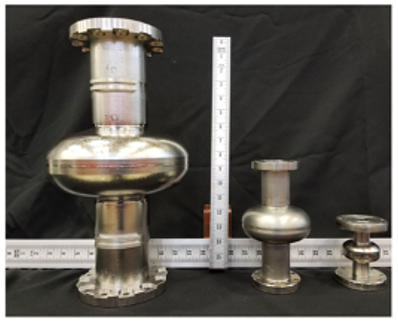}
  \caption{Single-cell TESLA shape SRF cavities. The diameter of the cavities scales as $\sim \frac{1}{f_r}$, where $f_r$ is the resonant frequency.}
  \label{fig:singlecell}
\end{figure}

Over the last decades, research in the field of particle accelerators (which use microwave cavities to propel the beam) has optimized the cavities material and fabrication techniques substantially, resulting into cavities with relaxations times about 2 seconds (internal quality factor $Q_I$ or $Q_0 > 10^{11}$) under the conditions required for quantum experiments \cite{romanenko2020}. An example of TESLA shaped cavities are shown in Fig. \ref{fig:singlecell}. This is several orders of magnitude larger than the best performing distributed resonators on a dielectric substrate (i.e. tantalum coplanar waveguide resonators on sapphire) for which $Q_I \sim 10^7-10^8$.

Given the ultra-high accelerating gradient and frequency stability obtained by the accelerator community, it is interesting to pursue the encoding of quantum information in similar cavities (colloquially referred to as superconducting radio-frequency (SRF) cavities), with very long relaxation time. This will however require introducing a non-linear element to the cavity, while avoiding additional losses. A common approach to achieve this involves coupling a non-linear superconducting oscillator, known as 'ancilla', to the cavity field. While such a coupling does make the cavity field weakly non-linear, the anharmonicity ($\alpha$, expressed by $(E_1 - E_0) - (E_2 - E_1)$) remains too weak to selectively drive transitions between cavities states. Over the years several techniques have been developed to bypass/alleviate this challenge without compromising the large Hilbert space offered by the harmonic oscillator.

\subsection{Encoding schemes}  
Several encoding schemes can be used for qudit operations on 3D QPUs. Some of them are currently being evaluated and analyzed. Advancing the selectively encoding of an ultra-high Hilbert space is partially in the scope of the SQMS center. More advanced implementations could go beyond the current mission of the center, and could lead to future R\&D activities.

\subsubsection*{Fock-basis encoding} 

Quantum information is encoded in the complex components of computational basis states. There are multiple proposals to represent information in the computational basis~\cite{binomial,gottesman2001encoding}. The simplest choice is a Fock state basis which represents a qudit as a superposition of photon number states. Also, a multi-qubit state can be mapped in a single qudit state on the Fock basis. For example, a three-qubit state can be mapped into a $2^3$ level qudit state. Then, counting the number of photons in the cavity is equivalent to reading out the states of each qubits. This qudit encoding scheme provides all-to-all connected qubits which are difficult to realize in the current 2D superconducting qubit devices.

Universal unitary control of a quantum harmonic oscillator can be achieved by coupling the oscillator with a non-linear ancilla element such as a qubit. In SRF cavity-based qudit architecture, a transmon is dispersively coupled to multiple modes in the cavity which enables universal control of each mode by the combination of cavity drives and transmon drives. There are two approaches to achieve universal control: numerically optimized control pulse technique and sequence of analytic gates such as (un)conditional qubit rotation gates and (un)conditional cavity displacement gates.

When a qubit is dispersively coupled to a cavity mode, the frequency of the qubit is split by the AC Stark effect, and each frequency corresponds to a photon number state of the cavity mode i.e.
\begin{equation}
\omega_q^{(n)} = \omega_q^{(0)} - n \chi,
\end{equation}

where the disperive shift is indicated by $\chi$. For low photon numbers, the magnitude of the dispersive shift is determined by the dipole coupling between the cavity and qubit ($g$) and the difference in resonance frequencies ($\Delta = \omega_q - \omega_C$) i.e. $\chi = g^2/\Delta$.

Driving the qubit in $\omega_q^{(n)}$ can be represented as $R^{(n)}(\theta,\phi) = R(\theta,\phi) \ket{n}\bra{n}$. This is called conditional qubit rotation gate which creates entanglement between the qubit and the cavity mode. When addressing a very large Hilbert space, we should also include the second order term in the photon number dependent shift approximation (Eq. \ref{secondorderchi}). $\chi'$ is usually orders of magnitude smaller than $\chi$, still over a certain $n$, this second order approximation leads to the indistinguishability between the qubit frequencies. This effect adds on to the dispersive shift approximation limit for high $n$. Therefore, to ensure the controllability of high Fock states' number, quantum computing devices need to be designed from the microwave point of view to achieve relatively low coupling and engineer $\chi$ to achieve large displacements ($\sim$\SI{}{\mega\hertz}) and fast controls.

\begin{equation}
\omega_q^{(n)} \approx \omega_q^{(0)} - (\chi n+ \frac{\chi'}{2} n^2) .   \label{secondorderchi}
\end{equation}

If a qubit evolves in a closed-loop trajectory in the Bloch sphere, it acquires a geometric phase that is proportional to the solid angle of the trajectory. Combining the property of a qubit with a conditional qubit rotation gate, it is possible to control the phase of the Fock state component of the cavity mode. This is called selective number-dependent arbitrary phase (SNAP) gate~\cite{PhysRevLett.115.137002}. The sequence of SNAP gates and unconditional cavity displacement gates enables a universal control of cavity state~\cite{krastanov2015}. The spectral bandwidth of the conditional qubit rotation gate pulse must be narrower than the qubit frequency splitting $\chi$ to achieve single cavity mode photon phase control in the SNAP gate. Therefore, the lower bound of the SNAP gate time is determined by $2\pi/\chi$, so the qubit coherence time should be substantially longer than $2\pi/\chi$ for a high-fidelity SNAP gate operation.

Despite the remarkable high relaxation times of the SRF cavities, the number of usable levels in the qudit encoding obviously remains finite. Practical limits to the Hilbert is part of an ongoing research. A somewhat optimistic boundary is defined by the minimum relaxation time required for an algorithm. Considering the relaxation time a Fock state scales inversely with the photon number, i.e. $T_1^{\ket{N}} = T_1^{\ket{0}} / n$, the highest Fock state is set by $\ket{n}_{max} =  T_1^{\ket{0}}/ T_1^{min}$. Assuming $T_1^{\ket{0}} = $ \SI{1}{\second} and $T_1^{min} = $ \SI{200}{\micro\second} the Hilbert space is limited to $\approx 5,000$ Fock states (equivalent to 12 qubits). In this estimation it is however implicitly assumed that the selective drive of Fock states remains feasible at these high photon numbers which is not a-priori correct. In fact, the exact diagonalization of the Jaynes-Cummings Hamiltonian of a single two-level system coupled to a cavity shows that the effective qubit-cavity coupling increases with photon number. Based on this, it is projected that well before a certain large photon number, known as critical photon number ($n_{crit} \approx (\Delta/2g)^2$), the dispersive regime is lost and the selective drive of the Fock states becomes impossible. Assuming $\Delta = $ \SI{2}{\giga\hertz} and $g = $ \SI{10}{\mega\hertz}, the critical photon number is about 10,000, which seems to indicate that a Hilbert space of 5,000 photons could still be operated. It should be noted that this would require a qubit with $\min [T_1, T_2]$ above 120 us, which is well within the range of current state-of-the-art transmon qubits. Experimental research has however shown both lower and higher critical photon numbers than would be expected. Moreover, in the estimate presented above, the Jaynes-Cumming Hamiltonian includes only a single two-level system and a single harmonic mode, which is a gross simplification for most experiments. Further research is required to identify the exact limitations to the maximum photon number. 

Fortunately, there are already several operational recipes known to bypass the critical photon number. A first example is the echoed conditional displacement (ECD) gate \cite{eickbusch2021} which requires a much lower qubit-qubit coupling than the Selective Number-dependent Arbitrary Phase (SNAP) gate and can be realized faster than $2\pi/\chi$ limit using large displacement in the intermediate stage of the gate operation. State preparation and control of a single qudit state in Fock basis encoding can be achieved by both SNAP gate and ECD gate protocols. Another interesting strategy uses selective Rabi drives in the presence of on-demand photon blockades. Beyond these examples it remains very relevant to focus ongoing and future research towards this topic to further extend the limits of the operational Hilbert space per cavity mode. 

\subsubsection*{Quantum Optimal Control (QOC)}
Together the SNAP and displacement gates form a universal gate set allowing for the creating and transformation of any arbitrary qudit state \cite{krastanov2015}. Expressed in terms of two-level systems, e.g. qubits, a harmonic oscillator of $N$ levels can encode and process the same about of information as $\log_2(N)$ two-level systems. This brings two notable advantages of the qudit encoding over e.g. interconnected qubits. First, the number of control lines required is reduced from $\log_2(N)$ to 1, which makes the system more scalable. Moreover, there is no need of special entangling gates to construct the full Hilbert space. Consider for example a collection of 4 two-level systems. To create the maximally entangled state, for instance $\ket{\psi} = (1/\sqrt{2})\left(\ket{0000} + \ket{1111}\right)$, a series of 3 consecutive entangling operations should be performed. Creating the corresponding state with a qudit, i.e. $\ket{\psi} = (1/\sqrt{2})\left(\ket{0} + \ket{16}\right)$, requires only a single operation. 

Advanced control of cavity states consisting of several consecutive SNAP and displacement gates can also often be further condensed into a single or few separate drive signals obtained by numerical optimization methods\cite{khanejaOptimalControlCoupled2005, defouquieresSecondOrderGradient2011, petersson2021optimal, gunther2021quandary, ozguler2022numerical}. This method has been successfully applied in various quantum systems \cite{doldeHighfidelitySpinEntanglement2014, andersonAccurateRobustUnitary2015} and most importantly was demonstrated to obtain universal gate sets for information processing in 3D cavity \cite{heeresImplementingUniversalGate2017}. Based on an accurate model Hamiltonian, a numerical optimization algorithm is used to construct the most efficient and accurate drive signal for the execution of a target unitary gate operation by varying several signal parameters (e.g. amplitude, phase, frequency) as a function of time. It has been shown that the gate duration for the SNAP gates engineered with optimal control signals are up to 8 times shorter than the analytical SNAP gate \cite{kudraRobustPreparationWignernegative2021}. Additionally, the model Hamiltonian used by the optimization algorithm can easily include higher-order terms which are often left out in analytical solutions, which significantly increases the gate-fidelity. 

While the optimization algorithms for medium scale systems can still be executed on a classical computer without significant complications, the size of the system Hilbert space is limited due to the amount of classical memory available. To some degree, this problem can be eased by allowing non-physical intermediate states during the optimization. Still, improving the performance of optimal control algorithms is the focus of current and future research efforts. 

\subsubsection*{Alternative encoding schemes}
Quantum information encoded in the Fock-bases is mostly affected by photon-loss events. Using ultra high-Q SRF cavities is a viable option to suppress this error-rate and can potentially lead to a hardware platform capable of executing algorithms with a reasonable (but finite) number of sequential gates. Yet, despite the progress, any further reduction of the relaxation time of SRF cavities will likely be challenging. Increasing the gate-depth any further, therefore requires either implementing an active error correcting scheme for qudits, or an alternative encoding scheme which is immune to photon-loss. Currently there are no active error correcting schemes known for qudits and future research should be focused to assess whether active error correction is feasible for a qudit encoding, see Section \ref{error_correction}.

The alternative, i.e. adjusting the encoding (or sometimes the Hamiltonian) to introduce protection against errors (e.g. photon loss) has been the subject to active research in the last decade and several different codes have been demonstrated successfully. The most straightforward example of these is the cat-code in which the information is encoded in a basis spanned by cat-states i.e. the superposition of two coherent states $\ket{\alpha}$ and $\ket{-\alpha}$ of a resonant mode, where $\alpha$ is a complex amplitude. The cat-bases typically used consists of 4 cat-states (non-normalized) i.e. $\ket{\mathcal{C}^{\pm}_{\alpha}} = \ket{\alpha} \pm \ket{-\alpha} $ and $\ket{\mathcal{C}^{\pm}_{i \alpha}} = \ket{i \alpha} \pm \ket{-i \alpha}$ such that any arbitrary superposition of the logical states $c_g\ket{g} + c_e\ket{e}$ can be expressed as $\ket{\psi_{\alpha}} = c_g\ket{\mathcal{C}^{+}_{\alpha}} + c_e\ket{\mathcal{C}^{+}_{i\alpha}}$. These basis-states are not completely orthogonal, but the wavefunction overlap decreases exponentially with $\alpha$ and already at $\alpha = 2$,  $|\braket{\alpha}{i\alpha}|^2 \ll 10^{-1}$. More importantly, the superposition-state is four-fold `cyclic' with respect to the photon number (i.e. losing 4 photon brings $\ket{\psi_{\alpha}} \rightarrow \ket{\psi_{\alpha '}}$ ) and photon-loss events can be 'counted' by tracking the state parity which makes correcting errors introduced by photon loss feasible~\cite{leghtas2013hardware}. While this demonstrates that a clever choice of encoding can introduce some error-protection, this encoding isn't truly convenient as, without any additional pumping, the photon number gradually decreases to zero. 

Protection against photon loss isn't specific to the coherent states (the bases of cat-sates) but can obtained as well by constructing a specific bases from Fock states. Codes based on this approach are referred to as binomial codes and have the advantage of consisting of a finite number of Fock states~\cite{binomial} as opposed to the broad and dense Fock state distribution required by coherent states. While the principle behind the protection, i.e. making the logical bases cyclic invariant to the photon number, is similar to the cat-based encoding, the original binomial codes require immediate interaction after a single photon loss. More recent and advanced work has shown that this issue can be resolved elegantly by engineering and actively driving the dominant photon-interaction between the cavity and environment. The technique, known as driven-dissipative stabilization, has been applied to both cat-codes \cite{Grimm2020a, Lescanne2020} and binomial codes~\cite{gertler2021protecting} and provides a very promising path to qubits with autonomous protection against photon loss. 

All these alternative codes come with a significant sacrifice considering only two logical states are exposed as opposed to the many states in a bare Fock-bases encoding, re-introducing all technical overhead as discussed earlier. The focus of future research in this field should therefore be focused on extending the ideas behind hardware-level protection to a qudit-based encoding (or visa-versa). 

\subsection{Scaling up to interconnected multi-mode cavities}
A single multi-mode cavity system potentially implements a Hilbert space large enough to run small quantum algorithms which puts this architecture well in the NISQ regime. Ultimately an even larger Hilbert space is required to run the more relevant quantum algorithms. A viable scaling strategy is therefore one of the most important parts of any quantum architecture. Adopting a different cavity geometry (e.g. the flute geometry \cite{vatsan_multimode}) in combination with a tunable coupler will certainly increase the number of modes coupled to the qubit, but becomes limited by the geometric coupling between the coupler and the qubit. Any further scaling will likely involve interconnecting different multi-mode cavities, each being controlled by a single qubit. The entanglement between the modes of different cavities can potentially be engineered by temporarily entangling the control qubits or perhaps directly via a tunable exchange interaction between the modes. Such an interaction has been demonstrated in Ref.~\cite{gao2019entanglement} between two single-mode cavities. Further research will have to show whether this approach can be extended (in a mode-selective manner) to two (or more) multi-mode cavities. 

\subsection{Multi-mode SRF cavities}

\begin{figure}[!ht]
    \centering
     \includegraphics[width=0.45\textwidth]{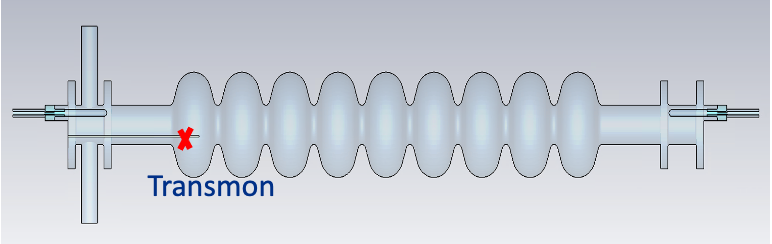}
  \caption{Microwave model of a multi-cell SRF cavity coupled to a transmon qubit.}
  \label{fig:multicell}
\end{figure}

Increasing the dimension of the Hilbert space is required to implement more complex algorithms. The best scaling architecture is still part of ongoing research in the community. As part of the SQMS efforts we are exploring the use of multi-cell cavities as multi-mode quantum systems.
Bulk Niobium (Nb) multi-cell cavities are routinely used in particle accelerators. These structures consist in the coupling of nominally equal single cells, resulting in closely spaced fundamental modes, all with very high quality factor ($Q_I \sim 10^{10}$), see Fig. \ref{fig:multicell}.

In this approach, which is similar to the work in \cite{vatsan_multimode}, a single qubit is dispersively coupled to a set of fundamental modes, each corresponding to a specific field distribution over the various cells of the cavity and acting as an independent harmonic oscillator. This quantum device is designed to achieve large relaxation time with equally sized Hilbert spaces. To make the spectral properties of the cavity better suited to encoding quantum information, the geometric design of the cavity can be optimized.  As an example, increased spacing between the fundamental frequencies can be achieved by acting on the cell-to-cell coupling; similarly, equal coupling for all the modes, can be achieved through tapering.

A major advantage of this architecture is the natural coupling of Fock states between the harmonic modes mediated by the non-linearity of the transmon qubit. Due to this, it is conceivable to engineer a generalized SNAP gate that operates on the combined Hilbert space of the multi-cell cavity. The goal of such a multi-mode conditional gate is to change the wavefunction of the entangled qubit-multimode cavity system only if certain photon numbers in the cavity modes are satisfied. In order to apply the SNAP gate on two cavities, for example, the qubit has to be driven with a photon dependent frequency, $\omega_d = \omega_q - n\chi_n - m\chi_m$, where $\omega_q$ is the qubit frequency, $\chi_{k} \approx g_k^2/(\omega_q - \omega_k)$ is the dispersive shift of mode $k = (n,m)$, $n$ ($m$) is the photon number in the respective cavity mode, $n$ ($m$), and $g_k$ is the dipole coupling between the qubit and the mode. Applying this gate will naturally lead to an entangled state of all the modes coupled to the qubit. Unfortunately, the always-on dispersive interaction between the qubit and all the cavity modes introduces a significant challenge here as the qubit drive is not just conditional on the photon occupation of two modes but on the exact Fock state superposition of each cavity mode. This challenge is part of ongoing research which will focus on the engineering of dynamic control of the dispersive shift of each individual mode and the qubit. For example, to perform a single cavity SNAP operation, a single dispersive shift will be tuned to a non-zero value, whereas a conditional two-mode SNAP operation can be implemented by setting the dispersive shifts of two modes to a finite value. The intuitive architecture that implements this kind of control consists of a multi-cell cavity (referred to as `storage') coupled to a single cell cavity (referred to as `control') via a tunable coupler. The control cavity is additionally coupled to a qubit which can be used to create and manipulate the Fock state in the control cavity. Before and after each state creation or gate operation, the Fock state in the control cavity can be swapped into one of the storage cavity modes using the tunable coupler which is either a tunable resonator or a non-linear resonator. In the first configuration the coupling is conveyed by 3-wave mixing process and will be selectively activated by modulating the resonator frequency at the difference-frequency between the single-mode cavity and the preferred mode of the multi-cell cavity. In the second configuration the coupling is conveyed by a 4-wave mixing process activated by driving the system with two tones such that $\omega_1 + \omega_a = \omega_2 + \omega_b$, where $\omega_{1/2}$ are the driving tones, $\omega_a$ is the single cell cavity frequency and $\omega_b$ is the specific mode frequency of the multi-cell cavity. Both configurations do however require more hardware and a careful engineering of the tunable coupler which urges for another, less hardware demanding solution to be found. Inter-connectivity schemes among quantum devices and memories, as well as tunable couplers are extensively treated in Section \ref{Interconnectivity}.

\subsection{3D QPUs to implement qudit algorithms for HEP}
One of the exciting potentials for circuit quantum electrodynamics (cQED) systems is the ability to implement these SRF platforms as quantum simulation devices for condensed matter systems (CMS) and systems. 
While there is a plethora of literature on quantum simulation on qubit hardware \cite{somma2003quantum, trabesinger2012quantum, ladd2010quantum, paulson2020towards, atas20212, Jordan2012, Banuls2020, georgescu2014quantum, macridin2021bosonic, roggero2020quantum, zache2021achieving, luo2020gauge, padmanabhan2014more, hashimoto2017time, bacon2006efficient,Davoudi:2022cah}, 
similar work on qudit based quantum platforms are rather unexplored territory. This is unfortunate, because for theories of interest to HEP, the number of local degrees of freedom can easily reach thousands and thus platforms with all-to-all connectivity highly desired~\cite{kan2021lattice,carena2022improved}.
In cQED platforms, the modes of the electric fields, or the Fock states are utilized as information encoding. 
As opposed to the standard $\{|g\rangle, |e\rangle\}$ levels on qubits, cQED systems can support more than 2 states, $\{|0\rangle, ..., |N-1\rangle\}$ with the Fock state encoding in one mode of the electric field.
We will call one of the $\{|0\rangle, ..., |N-1\rangle\}$ bosonic Fock states as qudit states. 
The goal of the simulation is to have a platform to carry out the simulations which are difficult to simulate on classical computers due to the exponential overhead of simulating quantum states.
The HEP simulations of scalar and gauge field theories on qudits have recently attracted interest \cite{klco2019digitization, klco2020systematically, 2021PhRvD.103i4501C,gustafson2021prospects,gustafson2022noise, 2021arXiv210813357M}.
In order to simulate the time-evolution of a field under a specific Hamiltonian, a digitization of the field must be made.  For scalar field theory this can be made with a Fock state basis.
First, the continuous scalar field $\phi(x)$ is digitized as $\phi_n$ \cite{klco2019digitization, macridin2021bosonic, 2021arXiv210813357M}. 
Then, this field can be expanded in a choice of orthogonal function basis and the harmonic oscillator eigenfunctions are discretized on the Fock states \cite{macridin2018digital, 2021arXiv210813357M}. In the case of gauge fields, digitization schemes exist in the Fock basis~\cite{2021PhRvD.103i4501C} or the group element basis~\cite{gustafson2021prospects,gustafson2022noise}. Using these formulations, the SQMS platforms could be used to simulate the dynamics of quantum field theories or predict quantities of interest in HEP like the viscosity~\cite{cohen2021quantum} or parton distributions~\cite{lamm2020parton} of low-dimensional quantum field theories.

The quantum simulations on qubits require application of 2$\times$2 Pauli gates and digitization of the field on qubit hardware. 
The unitary gates available on qudits are quite different from the usual Pauli gates we are familiar from qubit-based devices. 
One of the goals is to sketch out the available $SU(N)$ gates available on SRF devices at Fermilab.
The $SU(N)$ gates on SRF platforms are engineered with a control over the transmon and the electric field modes of the cavities. 
The control over the transmon and the electric field can be made via optimizing a time-dependent signal over the device. 
This approach is not always preferred, because the signal generator might have imperfections, and optimizing the signal on a classical computer is a computationally difficult task.
An alternative way of control over the cavity\&transmon system is to choose appropriate parameters for the gates available on the SRF platform. 
This can be done via variational methods although more investigation is needed when the problem size gets big.
Some of the available gates are SNAP, displacement gates, $SO(2)$ rotation gates with photon blockade, and SWAP gates to name a few. 
The displacement gate $D(\alpha)$ is defined as the following
\begin{eqnarray}
D(\alpha) = \exp\left[\alpha a - \alpha^* a^{\dagger}\right],
\end{eqnarray}
where $a$($a^{\dagger}$) annihilates (creates) a photon mode and $\alpha$ is a complex number. 
The diagonal SNAP gate on one $N$-level qudit is defined as the following:
\begin{eqnarray}
S_N(\Vec{\theta})   
=
\prod \limits_{n=0}^{N-1}e^{i\theta_n|n\rangle\langle n |},
\end{eqnarray}
where $\vec{\theta} = \left\{\theta_n\right\}$ is is a vector of occupation number dependent phases. 
Out of this `gate vocabulary' listed above, SNAP and displacement gates are particularly important due to their ease of implementation on hardware. 
It was shown in \cite{krastanov2015} that a combination of SNAP and displacement gates is universal. 
It is also possible to decompose unitaries on cQED systems using a sequence of Givens, or $SO(2)$ rotations, along with SNAP gates. 
The Givens rotation among two Fock states can be engineered by employing a photon blockade on a multimode cavity system. 
The time scale of the Rabi oscillation between two Fock states defines the SO(2) rotation angle, $\theta$ \cite{vatsan_multimode}. 
In theory, one hypothetical, `perfect' SNAP gate, which is a diagonal gate on the Fock state, that does not suffer from the infidelity can modify the local phases of an $N$-Fock state, thus this `perfect' gate can replace the subsequent applications of $2^d = N$ qubit gates, where $d$ is the number of equivalent qubits. 

The gates mentioned above are defined for a single cavity. 
Single cavities are insufficient for simulation because they offer limited number of Fock states in one qudit and controlling the connectivity in a single qudit is difficult. 
Thus, a full control over multiple cavities is required for a meaningful simulation. 
In theory, a straightforward engineering of multiqudit SNAP gate, or generalized SNAP gate, could be useful for the simulations. 
To be more concrete, an arbitrary SNAP gate acting only on the qudit $k = 3$ among 5 qudits should look as 
$
S(\Vec{\theta}_k)= \textbf{1}_N\otimes\textbf{1}_N\otimes S_N(\Vec{\theta})\otimes\textbf{1}_N\otimes\textbf{1}_N
$,
where $\textbf{1}_N$ is the $N\times N$ identity matrix and $S_N(\Vec{\theta})$ is a single qudit SNAP gate on qudit number $k$ only.
The Trotter simulations on a qudit is straightforward with the multiqudit SNAP gates. 

Another useful gate for quantum simulation is the echoed displacement gate. 
The echoed displacement gate can couple the degrees of freedom of the transmon to the bosonic Fock states in the oscillator,  and so one could imagine simulating, via for example a Jordan-Wigner transformation, a fermion system coupled to a bosonic system.
One important requirement for many quantum simulation algorithms is the ability to implement general-group quantum Fourier transforms. For example, in lattice gauge theories, a group Fourier gate is required for kinetic energy Trotter step \cite{lamm2019general,alam2021quantum,gustafson2021prospects,gustafson2022noise}. For $D_N$ gauge theory simulations, the non-Abeian Fourier gate which diagonalizes the kinetic energy gate has been explicitly constructed \cite{alam2021quantum}.
The SWAP gate based on multiphoton blockade is useful for HEP simulations on a discretized lattice. SWAP gate is formally defined as swapping the phase of one Fock state. To be concrete, SWAP operation on a three-Fock state basis corresponds to the operation of 
$
|\psi\rangle = (1/\sqrt{3})\left(|0\rangle + e^{-i\phi}|1\rangle + |2\rangle\right) \rightarrow (1/\sqrt{3})\left(|0\rangle + |1\rangle + e^{-i\phi}|2\rangle\right)
$. This operation can also be done with employing the multiphoton blockade between two photon levels \cite{chakram2020multimode}. 

One of the hardware difficulties when implementing a quantum simulation algorithm is the time scale of the transmon and cavity. Any unitary gate acting on the qudit for simulation purposes should be repeated many times. The total time required for the gates should be within the decoherence times of the cavity, whereas the duration of each single gate should be well within the decoherence times of the ancilla qubit.

\subsection{Materials for 3D QPU}
Clearly the performance of the 3D quantum architecture benefits from higher coherence times of both the cavity (larger gate depth) and the ancilla qubit (longer gates and consequently a Hilbert space). One promising path towards coherence time improvement is the in-depth investigation of materials losses in both cavities and qubits. Here we focus on the material losses in bulk cavities. In section \ref{2d_materials}, we will focus on the material losses in planer quantum devices. 

Niobium represents one of the most used metal for SRF cavities. The performance of these cavities depends greatly on a couple of key factors such as the presence of a niobium pentoxide that may form on the surface under ambient conditions as well as the purity of the niobium. The niobium pentoxide, in particular, can host two-level system defects which in turn introduces significant quality factor degradation \cite{PhysRevLett.119.264801}. To this end, a variety of purification techniques have been developed to improve the quality of niobium cavities. These include targeted heat treatments, where the cavities are heated for several hours to \SI{300}{\celsius} - \SI{400}{\celsius} under vacuum, to remove the surface oxide \cite{romanenko2020}.

Through these developments, niobium cavities have recently demonstrated photon lifetimes up to 2 seconds along with quality factors exceeding 10$^{10}$ in the quantum regime \cite{romanenko2020}. This 1000-fold increase in coherence times makes implementing these cavities promising for 3D QED architecture. Further increases in photon lifetimes necessitate an improved understanding of the relationship between additional impurities in the niobium metal, such as hydrogen and carbon, and cavity performance. Studies related to these topics as well as mitigation strategies to boost performance are currently underway.

It is important to understanding how the material properties of superconductors, dielectrics, oxides, and related interfaces relate to the performance of the quantum devices. The relative contributions of various regions (superconductor vs dielectric substrate vs surface oxides) to the total loss of a device in tandem with investigations of the microscopic structure of TLS defects allow the optimization of quantum device coherence, and thus are key factors for the realization of large-scale quantum computing architectures.

To this end, in parallel with structural characterization of quantum devices' materials, several efforts at SQMS are focused on studies of internal quality factors of 2D and 3D resonators. 2D Superconducting resonators are constituent parts of many currently explored quantum architectures and they suffer from the same loss mechanisms as superconducting qubits, while being conducive to measurements that allow the distinction between loss channels \cite{mcrae2020materials}. It follows then that understanding and mitigating losses in resonators can guide improvement of qubit performance. Characterization of 2D and 3D resonators is performed in relation to various fabrication and processing techniques as well as interface engineering methods aimed to improve the material properties of superconducting quantum devices and, ultimately, device performance.

One of the efforts towards a greater understanding of microscopic TLS behavior is focused on the characterization of the spectral line widths of TLS defects in Nb SRF cavities by means of two-tone spectroscopy. In this setup, one of the fundamental pass band modes of a multi-cell SRF cavity is pumped with the high power signal to saturate the TLS defects, and the internal quality factors of other modes are probed with signals of much smaller amplitude. This project could provide a way to distinguish between different types of TLS defects using their spectral fingerprints and is a first steps towards the demonstration of a novel defect spectroscopy technique. 

\section{2D QPU}\label{2DQPU} 
The 2D superconducting circuit is currently the most widely and successfully implemented architecture for quantum computing. On this platform, especially with the implementation of transmon qubits numerous research labs and a number of well-known companies, including Google, IBM and Rigetti, have made impressive achievements toward practical quantum computers. Notably, the first experiments claiming the quantum advantage over classical machines have been carried out on such platform. Compared with 3D devices, 2D chip-based systems enjoy mature fabrication and microwave control techniques. It is much more convenient to introduce various control knobs on the 2D platform, including charge and flux control lines, which are crucial for high-fidelity quantum operations.

The superconducting qubit Round Robin experiment at SQMS investigates, reveal, and distinguish between environmental decoherence channels, including cosmic ray-induced quasiparticle generation, in addition to providing cross-institutional benchmarking for SQMS superconducting platforms using Rigetti multiqubit chips. This first-of-its-kind study will provide new insights toward advancing state of the art performance, as well as allowing high-precision cryogenic microwave materials investigations. It has successfully progressed at all involved institutions including FNAL, NIST, INFN, and Colorado Boulder. FNAL is also pursuing advanced fabrication technologies and optimized device designs to minimize the two-level defects, mitigate non-equilibrium quasiparticles, and suppress charge/flux noises, all of which are aimed to improve the coherence performance of single qubits and multi-qubit up-scale capability. The superconducting quantum circuits are fabricated at three state-of-the-art cleanroom foundries at Rigetti, NIST, and U. Chicago. The device characterization will be performed at FNAL, Rigetti, INFN, and Colorado Boulder in multiple quantum testbeds at sub-mK temperature range. 

\subsection{Qutrit implementations}
The transmon is a weakly anharmonic oscillator, therefore energy levels higher than the lowest two levels can be utilized for quantum computation. Extra energy levels will enable larger Hilbert space. The coherence of higher energy levels was investigated thoroughly in Ref. \cite{petererCoherenceDecayHigher2015}. However, climbing up this energy ladder comes at the cost of coherence time, which is in fact reduced. Explained by Fermi golden rule, the coherence time of higher energy levels will decrease linearly. Namely, the second excited state will have only half of coherent time compared to that of the first excited state. In addition, charge noise will also become significant. Therefore, up to now, only the lowest three levels of the transmon which create a qutrit have been studied and used in several application, in particular for 2D platforms. Despite the coherence limitation, qutrit has been able to offer various advantages. The third energy level was used to improve the qubit operation, namely shortening three-qubit gate such as Toffoli gate, CCZ gate \cite{fedorovImplementationToffoliGate2012, galdaImplementingTernaryDecomposition2021, hillRealizationArbitraryDoublycontrolled2021}, error detection \cite{rosenblumFaulttolerantDetectionQuantum2018}, readout fidelity enhancement \cite{malletSingleshotQubitReadout2009} and even enable quantum state transfer for quantum communication \cite{kurpiersDeterministicQuantumState2018}. In terms of error correction, the theory has clearly shown advantages such as superior error thresholds and yield in magic-state distillation \cite{campbellBoundStatesMagic2010, campbellMagicStateDistillationAll2012}. Furthermore, the complexity of certain algorithms will be reduced greatly \cite{gokhale2019}. Finally, there has been theory work for noise reduction when using qutrits for quantum simulation of gauge theory which will be discussed in more details in the qutrits algorithms for HEP. From the implementation point of view, qutrits approach will utilize current 2D qubit infrastructure (provided by Google, IBM, Rigetti) or require minimal changes. Rigetti already has qutrits operation support, IBM allows pulse level control which can be used for qutrits operation by users.

Over the past two decades, the coherence of 2D superconducting qubits has been boosted over five orders of magnitude. This is realized by both reducing the noise in the device and suppressing the coupling between the qubit and the noise. However, even with this excellent improvement, the state-of-the-art coherence is on the order of hundreds of microseconds. 
On one hand, the lifetime of the qubit is currently limited by high-frequency noise, such as dielectric loss and quasiparticle loss. On the other hand, the coherence of the qubit is affected by low-frequency noise, such as $1/f$ charge noise and flux noise. 
The noise channel of 2D resonators is also simple, i.e., only limited by photon loss. Since the resonator frequency is determined by the geometry, which is rather stable, the intrinsic pure dephasing is negligible. This asymmetry in bit-flip and phase-flip error, known as the biased noise, also shows certain advantage in the context of quantum error correction. A challenge for 2D qubits is the ability to fabricate large multi-qubit systems with long single-qubit coherence and precise parameter targeting. While 3D QPUs have intrinsic all-to-all connectivity, additional wiring is required to realize equivalent architectures in 2D QPUs, which can be complicated in practice, and may introduce additional noise channel. 
2D planar quantum circuits with 256 physical qubit processor and long coherence times, approaching milliseconds, would enable two-qubit gate error rates below thresholds for fault-tolerance and bringing quantum computers entering the quantum advantage era.

\subsubsection*{Qutrits algorithms  for HEP} 
Recently, qutrit formulations for problems in HEP have begun gaining traction \cite{gustafson2021prospects,gustafson2022noise,2021PhRvD.103i4501C, gustafson2022noise,2021arXiv210813357M,2021PhRvD.103i4501C}. As discussed above, one advantage in using qutrits over qubits is the reduction in entangling gate requirements. Another benefit of qutrit-based devices for HEP is that the local degrees of freedom often are not powers of 2, e.g. for many truncations of $U(1)$ \cite{2021PhRvD.103i4501C, gustafson2022noise, Unmuth_Yockey_2018, Unmuth_Yockey_2019, Bazavov_2015, PhysRevLett.121.223201} and for crystal-like subgroups of $SU(2)$ and $SU(3)$ \cite{Alexandru_2019,Ji_2020,ji2022gluon,alexandru2021spectrum}. When mapping these theories onto qubits, this mismatch leads to qubit states which do not correspond to any state in the field theory, which is both inefficient and without error correction can lead to large errors in simulations. In contrast, mapping these theories onto qudits or qutrits can avoid or reduce the number of such mismatched states. Work such as in Refs. \cite{2021PhRvD.103i4501C, gustafson2022noise} implement algorithms assuming a gate set of arbitrary 1-qudit rotations and a controlled increment gate which is defined as 
\begin{equation}
    C_{inc.} = \sum_{i = 0}^{N - 1} \sum_{j = 0}^{N - 1} |i\rangle\langle i| \otimes |j\rangle\langle (j + i)_{mod N}|.
\end{equation}
A tentative noise study on qutrits versus qubits indicates that even if qutrit gates are significantly noisier than qubits, equivalent physics can be achieved \cite{gustafson2022noise}. It is open study on how these results may extend to higher number states and other theories.

\subsection{Many-body correlation \& scrambling detection and simulations} 

With recent developments in the increase of quality of building blocks of quantum circuits, we are moving from physics demonstration into physics discovery era. Many-body systems constitute useful problems that can be hard to simulate in supercomputers and can also be used as benchmarks to test performance of the quantum simulators \cite{blok2021quantum, mi2021information, ozguler2021excitation, chen2022error}. Recent studies show that quantum simulator performance has almost caught up the performance of supercomputers for certain problem instances \cite{blok2021quantum, mi2021information}. Qudit-based processors will foster physics discoveries. In addition to providing lower-depth circuits and noise improvement with hardware-efficient solutions \cite{wang2020qudits, gustafson2021prospects, gustafson2022noise, otten2021impacts}, qudits with even small number of levels can provide rich structures \cite{wang2018qutrit, blok2021quantum}. Each mode in bosonic quantum processors can host many levels, which are natural candidates for qudit simulations \cite{ozguler2022numerical}.

Quantum circuits can be used as platforms to simulate spreading of information, which include but not limited to optimization problems encoded in spin models \cite{lucas2014ising, hormozi2017nonstoquastic, ozguler2018steering, cao2021speedup, li2021benchmarking}, unconventional many-body phenomena, such as many-body localization and scars \cite{basko2006metal, turner2018weak, ozguler2020response, iadecola2020quantum} and lattice quantum chromodynamics \cite{wiese2014towards}. An information theoretical aspect that merges seemingly different scales under one roof, from spin glass to black holes, is scrambling, which is defined as the spreading of localized information from a small part of a system to the entire system. The exponentially fast spreading is a signature of quantum chaos. Squared commutators and out-of-time-order correlator (OTOC) are used as measures for the scrambling and the total accessible degrees of freedom in the system \cite{maldacena2016bound, huang2017out, roberts2017chaos, swingle2017slow, rozenbaum2017lyapunov}. Spreading of operator entanglement is hard to simulate classically but it can be measured efficiently by OTOC fluctuations on the quantum device \cite{mi2021information}. Squared commutators of black holes grow at the fastest scale possible (they are "fast scramblers" \cite{sekino2008fast}). AdS/CFT correspondence can be used to design unitary actions describing black hole dynamics, which provides new perspectives to study quantum gravity in the lab \cite{brown2019quantum, nezami2021quantum}.

\subsection{Materials for 2D QPU}\label{2d_materials}

Similar materials considerations to those discussed for 3D QPUs are relevant in the case of 2D QPUs. Niobium metal once again represents one of the most employed superconducting material. In the case of 2D QPUs niobium is deposited as a thin film on an underlying dielectric substrate to attain a 2D geometry. This introduces the presence of various interfaces and surfaces which can limit coherence times and other performance metrics. These include factors such as the interface between the metal thin film and the underlying substrate (metal/substrate interface), interfaces between differently oriented metal grains in the niobium film (grain boundaries), the surface oxide that may form on the unexposed substrate (substrate/air interface), and the niobium pentoxide that forms at the surface of the niobium metal discussed previously (niobium/air interface) \cite{Nersisyan2019, murthy2022potential}. Each of these can introduce loss in the 2D system. Further, lithography processes used to define the superconducting circuit can also introduce interstitial impurities in the niobium metal that can negatively impact the superconducting properties \cite{doi:10.1063/5.0079321}. As such, identifying schemes to mitigate the impact of these factors remains an active area of materials research.

\begin{figure}[!ht]
    \centering

     \includegraphics[width=0.4\textwidth]{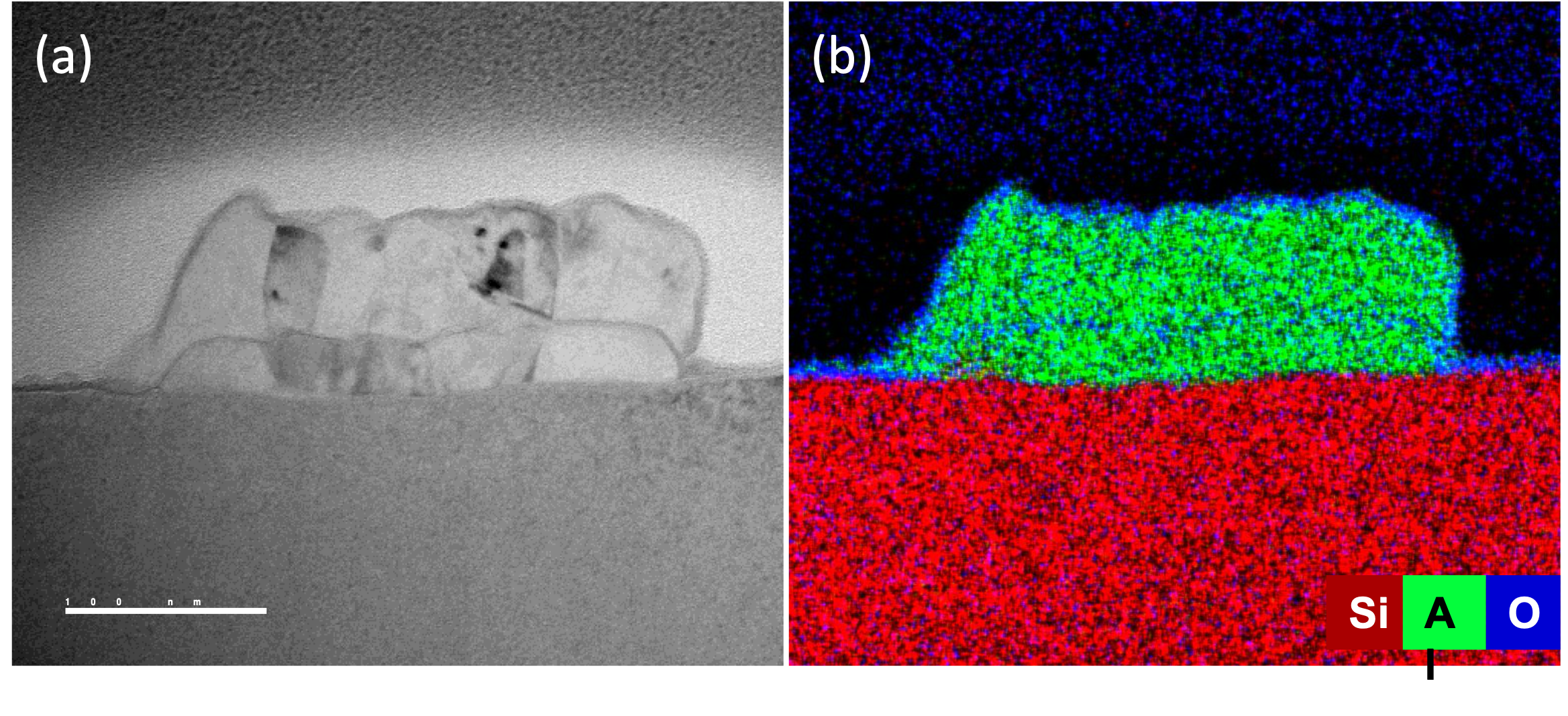}
  \caption{ (a) Bright -field transmission electron microscopy image and (b) corresponding Energy Dispersive X-ray Spectroscopy (EDS) elemental mapping of a Josephson junction made by aluminum metals with ultra-thin aluminum oxide in between.}
  
  \label{fig:EDS_AMES}
\end{figure}

Further, aluminum metal is traditionally used for the two superconducting electrodes in the Josephson junctions, which are used to introduce anharmonicity in these systems. These electrodes are separated by an ultra-thin insulating barrier through which electrons can quantum mechanically tunnel. The most common method for fabricating this tunneling barrier involves a diffusion-limited oxidation process of aluminum, which leads to the formation of an aluminum/aluminum oxide/aluminum junction, as shown in Fig. \ref{fig:EDS_AMES}. In terms of loss mechanisms, spatial variations in the oxide barrier width can lead to inhomogeneity in the tunneling currents and induce variability in operating frequency of the qubit. Further, similarly to niobium pentoxide, aluminum oxide can host two-level system defects and impact coherence times in these systems \cite{towards2019}. To this end, the identification of new methodologies for fabricating higher quality Josephson junction free of these loss mechanisms has garnered significant research interest.

\section{Inter-connectivity}\label{Interconnectivity} 
Scalable quantum computing, which is one of the ultimate goals of the SQMS center, must address the question of distributed quantum information processed among spatially separated nodes with high fidelity and high efficiency. Interconnecting superconducting quantum devices in distributed quantum networks would expand the capabilities of current superconducting QPUs, enabling operations, entanglement and state transfers beyond a single cryostat.  
In quantum communication, this would enable securing communication channels, e.g. through key distribution cryptography protocols. Moreover, networks of quantum sensors could enhance the measurement sensitivity below the standard quantum limit (SQL), or enable new sensing experiments and schemes \cite{awschalom2021development, Brady:2022bus}.  

At ETH Zurich, quantum teleportation was demonstrated connecting two dilution refrigerators through a cryogenically cold \SI{5}{\meter}-long microwave line \cite{magnard2020microwave}. There are advantages in using the superconducting technology for both quantum communication and quantum processing, such as high fidelity and error correctable chip-to-chip communication. However, this concept is not suitable for long-distance communication, which is essential for creating larger quantum networks or expanding the energy range of quantum sensors. In fact, while superconducting quantum systems can process quantum information in a fast way at milli-Kelvin temperatures, optical photons are natural information carriers over long distances at ambient temperatures. 

In this section, we discuss the technical challenges in the implementation of quantum networks through interconnected nodes and hybrid transduction devices. Several protocols can be considered for quantum communication, some of them may include the use of variable couplers to synchronize catch and release operations. We also describe different tunable schemes for qubits coupled with SRF cavities and 2D QPUs connected to 3D quantum memories.

\subsection{2D QPU and 3D quantum memory} \label{sec:2DQPU} 
Although the coherence time of 2D QPU is orders of magnitude smaller than the 3D one, it has the advantage of faster gate speed. To exploit the distinct features from different computing components, it can be promising to build a hybrid QPU for some applications. More specifically, we store the information in the 3D cavity to take advantage of its long coherence time. When performing gates, we transfer the quantum state to a 2D QPU which then enables fast gate operations. 
This hybrid QPU approach can also be generalized and actually has a close analogy to the error protected qubit introduced in Sec. \ref{error_correction}. In general, a computational subspace with long coherence time means that it is protected against the noise environment. However, this also makes it harder to manipulate the information, i.e., to control the device. Therefore, protocols to temporarily leave the protected subspace to perform a gate and then map back after the gate is finished are needed.

\subsection{Interconnection between quantum computation nodes} 
Physically, distributing quantum information processed among spatially separated nodes amounts to transporting quantum states between nodes via photon channels, a process often termed Quantum State Transfer (QST). Large entanglement rates between nodes are key to distributed fault-tolerant quantum computation \cite{YC_Jiang2007}. The methods to achieve this, specific to SQMS’s unique architecture, will be a logical next step in R\&D following matured high-fidelity manipulation of individual nodes. We envision a staged approach with milestones clearly identified, building up expertise first in well-established techniques adapted to the SQMS environment, while eventually aiming for cutting-edge studies with breakthrough potential.
In its essence, a QST task aims to accomplish the following:
\begin{equation}\label{eq:QST}
\begin{split}
\left(\alpha\ket{0}_1+\beta\ket{1}_1\right)\otimes\ket{0}_P\otimes\ket{0}_2 \xrightarrow{\text{QST}} 
\\ \longrightarrow\ket{\epsilon}_1\otimes\ket{\delta}_P\otimes\left(\alpha\ket{0}_2+\beta\ket{1}_2\right),
\end{split}
\end{equation}
where the initial quantum state in the product space of node 1, node 2, and photon channel $P$, with node 1 in a preset pure state and node 2 and channel $P$ in the ground and vacuum states, is mapped into the final quantum state in which node 2 becomes a copy of the initial state of node 1 and node 1 and channel $P$ are left in potentially indeterminate states.

Left to passive evolution by the Hamiltonian, this goal cannot be achieved unless the coupling between the nodes and the channel can be programmatically varied in time \cite{YC_Cirac1997,YC_Jahne2007}.  The potential afforded by this additional handle was behind most of the developments in QST \cite{YC_Korotkov2011,YC_Stannigel2011} since its first proof of principle.  The main challenges to a particular QST scheme can be summarized as follows:
\begin{enumerate}
    \item 	{\it Performance.} The overriding criterion of QST is the overall fidelity, $F= \lvert \bra{\psi_1}  \ket{\psi_2} \rvert ^2$, where $\ket{\psi_1}$ and $\ket{\psi_2}$ are the quantum states of initial node 1 and final node 2.  This fidelity should meet the demands of a massively scalable quantum computer.
    \item {\it Efficiency.} Gate time and error tolerance requirements translate into parameters of the QST coupling modulation, such as total transmission time and the on/off ratio measuring the dynamic range of the coupling strength \cite{YC_Korotkov2011}. 
    \item 	{\it Robustness.} QST is susceptible to a wide range of systematic and dynamic sources of errors \cite{YC_Sete2015}, not the least detuning between the emitting and receiving nodes, with $\sqrt{1-\eta} \propto \abs{\delta \omega}$ , where $\eta$ measures the QST fidelity and $\delta \omega$ is the detuning.  Detuning, either static or modulation induced, must be controlled to tolerable level. 
     \item {\it Environment.} Success of the QST scheme depends on the initial photon channel being in the vacuum state, as well as effective suppression of thermal photons to an exacting level.  This condition will be ensured initially with the entire system in the milli-Kelvin environment \cite{YC_Kurpiers2017}.  Proposed concepts in Refs. \cite{YC_Xiang2017, YC_Vermersch2017, YC_Leung2019} can potentially be adopted or expanded to lift this restriction with photon channels thermalized up to a few Kelvin, holding promise for realistic  quantum computing distributed over nontrivial distance.
    \item {\it Control configuration and algorithm.}
a.	Coupling modulation scheme – Of key importance is the decision on the method for coupling modulation, either magnetic \cite{YC_Bialczak2011, YC_Yin2013, YC_Wenner2014}, microwave driven \cite{YC_Zeytinoglu2015, YC_Pechal2016}, or other.  This determines how the emitted/absorbed photon wave function is controlled.  
b.	 Waveform implementation – The coupling modulation waveform is not unique, nor is the digital control hardware/software implementation.  Major study is required to arrive at optimized plans.
Our project thus amounts to systematic resolution of the above challenges.  We propose to divide the entire project into two stages.  The first stage aims at demonstrating high fidelity QST with good efficiency (total time, on/off ratio, etc.) on the SQMS-specific architecture in a milli-Kelvin environment.  Key components of this stage are listed below.  Breakdown into detailed tasks involving design, fabrication, algorithm, procedure, protocol, etc. shall precede implementation of the following components. 

\begin{enumerate}
\item	Hardware and infrastructure making up the QST topology.  This includes an emitting node, which is a resonator with or without a qubit capable of holding user-specified states and coupled to the transport channel, a receiving node with similar characteristics, and the transport channel. In view of the high-Q 3D cavity dragging out process time and imposing exacting tolerance on detuning between the emitting and receiving nodes. The hybrid option discussed in section \ref{sec:2DQPU}, with 2D QPU acting as an intermediary, may provide a viable mitigation to these challenges.
\item 	Coupling mechanism between the emitter/receiver and the channel, and provision for modulating the coupling strength \cite{YC_Korotkov2011, YC_Stannigel2011, YC_Bialczak2011, YC_Yin2013, YC_Wenner2014, YC_Zeytinoglu2015, YC_Pechal2016}.  This requires comprehensive design study in terms of nature of coupling modulation (magnetic or microwave driven) as well as optimal parameters.  The decision on coupling and modulation scheme will be the first item on the agenda.
\item	Choice of coupling modulation mechanism has implication on the next level staging.   QST demo between resonators alone as an intermediate step may be possible via magnetic tuning \cite{YC_Bialczak2011, YC_Yin2013, YC_Wenner2014}, whereas it is not obvious how to explore microwave-based tuning  \cite{YC_Zeytinoglu2015, YC_Pechal2016} without qubits in the resonator.
\item	Waveform of coupling modulation and software/hardware provisions needed to realize such modulation.
\end{enumerate}
All the above tasks will leverage Fermilab’s expertise in fabricating and operating superconducting infrastructure and hardware, microwave electronics and control, and numerical modeling/simulation.
Planning for the second stage should be carried out concurrent with the above tasks, aiming for QST demo over channels above the milli-Kelvin level \cite{YC_Xiang2017, YC_Vermersch2017, YC_Leung2019}, thus increasingly susceptible to thermalization of the quantum state during transport. This holds great implication to scalable quantum computation in the real world. Several published concepts will provide insights and inspirations for this study. It is nonetheless conceivable, as an initial attempt, to explore enlarging the parameter envelope through trade-off between transmission channel temperature and parameters already achieved in the first stage. This effort alone would produce valuable data and insight into the global problem.
In terms of infrastructure, the second stage will call for introducing an above milli-Kelvin transport channel, possibly in stages, into the baseline topology, with attendant modifications to be identified. 
\end{enumerate}    
    
\subsubsection{Tuning of coupling strength}

From the pioneering work of Cirac et al. in Ref. \cite{YC_Cirac1997}, it was recognized that in order to achieve fidelity arbitrarily close to 100\% in QST, the waveform of the transported photon must be tailored through modulating the coupling strength, $\kappa(t)=\sqrt{\gamma(t)/2\pi}$.  Further works in Refs. \cite{YC_Jahne2007, YC_Korotkov2011} refined on this scheme, leading to time-symmetric pulse shaping on both ends of the transmission channel, effectively causing the wave functions of the reflected photon and back-transmitted photon to form destructive interference at the transmission channel-receiver boundary, thus making possible the perfect transmission from emitting to receiving nodes.  There is certain freedom in choosing the detailed pulse shape \cite{YC_Stannigel2011}, whereas extensive work has been done on mechanisms to realize pulse shaping.  Mechanisms to be explored include magnetic flux modulated coupling \cite{YC_Bialczak2011,YC_Yin2013, YC_Wenner2014}, microwave induced qubit-resonator coupling, and potentially other methods \cite{YC_Zeytinoglu2015,YC_Pechal2016}.

\begin{figure}[htbp]
    \centering
     \includegraphics[clip, trim=0cm 20cm 0cm 0cm, width=0.45\textwidth]{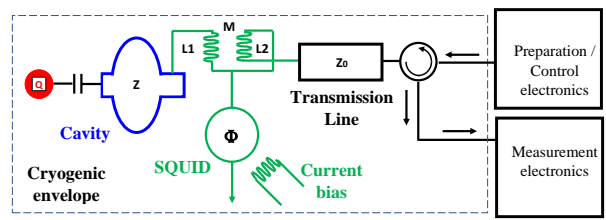}
  \caption{Schematic of node-to-transmission channel coupling modulation via external flux control. The resonator (blue) is capacitively coupled to a qubit (red) on one end and coupled to the transmission channel (black Z$_{0}$) through a tunable inductive coupler.  Coupler control is through the variable inductance (SQUID) by the bias current.  Control, preparation, and measurement electronics are on the right side.}
  \label{fig:node transm2}
\end{figure}

\paragraph{Tuning through magnetic inductance.}
A typical configuration consists of the node and the transmission channel coupled through a SQUID based tunable coupler.  The coupler is made up of fixed inductances with mutual inductance, and a flux tunable SQUID.  As the flux through the SQUID is varied by a bias current, the total effective inductance, and thus the resonator-transmission channel coupling, is modulated.  To implement a complete node system, as well as to be able to initialize the node in a particular quantum states, the node will consist of a qubit embedded inside a resonator.  Figure \ref{fig:node transm2} depicts this configuration.

As this implementation calls for design and hardware, especially SQUID based coupling to 3D cavities, which are new areas of research, challenges may arise.  For example, with a 3D resonator, the flux coupling geometry may compromise qubit lifetime \cite{YC_Reed2013}.  These potential concerns would warrant careful analysis/modeling and/or prototyping.  

\paragraph{Microwave-induced tunable coupling of a Raman process.}

This is a technique that achieves effective tunable coupling between a resonator-qubit node and a transmission channel, thus pulse-shaped quantum state transfer \cite{YC_Zeytinoglu2015, YC_Pechal2016}.  It takes advantage of a cavity-assisted Raman process where, due to anharmonicity in the qubit, a coherent microwave drive near the frequency gap between the dressed $\ket{f,0}_D$  and $\ket{g,1}_D$ states with waveform $\Omega(t)$ will create, in the lowest order of perturbation, an effective Jaynes-Cummings coupling between these two states.  Here $\ket{g}, \ket{e}, \ket{f}, \ket{0}, \ket{1}$ denote the ground, 1st, and 2nd states of the qubit, and the first two states of the cavity.  The new, time varying, coupling in the effective perturbative Hamiltonian is:
\begin{equation}
    \tilde{g}(t)=\frac{g\Omega(t)\alpha}{\sqrt{2}\Delta(\Delta+\alpha)},
\end{equation}
where $\alpha$ is the qubit anharmonicity, $g$ the qubit-resonator coupling, and $\Delta$ the qubit-resonator detuning.  This tunable coupling provides a handle for controlling the transition rate between the  $\ket{f,0}_D$ and the $\ket{g,1}_D$ state, with the latter coupled through decay rate $\kappa$ to global ground state $\ket{g,0}$ by emitting one photon into the transmission channel.  So overall a tunable mechanism responsible for the emission of a photon wavefunction into the transmission channel is obtained in terms of the combination of $\tilde{g}(t)$ and $\kappa$.
A self-consistent solution to the modulation function can be further derived, in the simple case of only two levels (i.e., states, e.g., $\ket{f,0}_D$  and $\ket{g,1}_D$) in the manifold corresponding to the effective Jaynes-Cummings term, that transmits one complete photon into the channel:
\begin{equation}
    \tilde{g}(t)=\frac{\kappa}{2\cosh{(\kappa t/2)}}.
\end{equation}

In both tunable coupling schemes, the waveforms of emitted and absorbed pulses are not unique.  They however display time-reversal symmetry.  This, as well as the stringent tolerance on cavity detuning, reflects the exquisite demand on precision for the success of this task.

\subsubsection{Flux-tunable qubits in SRF cavities} 
Several novel tunable superconducting qubit designs have been proposed over the past decade. In this list are flux, fluxonium and $0-\pi$ qubit designs which promise decreased sensitivity to external noise and perturbations and, consequently, improved coherence times ~\cite{yan2016flux, HCFluxonium, Brooks2013}. One common feature of these devices is that their control and operation rely on externally applied magnetic fields that are needed to reach the optimal regime.

Delivering external magnetic fields inside a 3D cavity while retaining the advantage of low loss of 3D SRF architectures is a non-trivial challenge. Even more so if one aims to reach the addressability of individual qubits inside a cavity.

Using standard approaches such as flux lines or external magnetic coils is challenging with 3D cavities. In the former case, superconducting flux lines are prone to coupling to the resonance modes of a cavity leading to unwanted parasitic interaction between the qubits and the cavity modes as well as distortion of the cavity resonance mode fields leading to additional losses. In case of externally applied magnetic field, it is difficult to completely avoid magnetic flux trapping in the walls of a cavity leading to losses associated with the motion of superconducting vortices. In the latter scheme, the individual addressability and tuning of qubits will not be possible once quantum circuits reach scales beyond several qubits.

An alternative approach can be to use micro- and nanomagnets to locally generate magnetic fields needed to tune and control the qubits inside an SRF cavity. Recent advances in all-optical switching (AOS) of magnetization on femtosecond time-scales in thin films and nanomagnets ~\cite{PRL2007AOS, SinglePulse_AOS2020} provide a viable approach to control magnetic flux sensitive qubits without usage of superconducting current-caring lines or application of external magnetic fields disadvantages of which are discussed above. In case of AOS, magnetization of small areas of magnetic films (few {$\mu$m} to few tens {$\mu$m}) is controlled by means of short laser pulses that toggle magnetization between two stable orientations (up or down) which are typically perpendicular to the plane of a film. The size of the switched area is defined by the width of the laser beam and magnetic properties of the film. Both parameters are readily controlled by tuning the optical system and dimensions of the optical waveguides and the magnetic film growth and patterning parameters set during the fabrication process. In general, optically controlled micro- and nanomagnets present many parameters that can be engineered to desired values such as total size (including thickness), magnetic moment, magnetic anisotropy, material composition, optical pulse intensity and duration, which is need to switch the magnetization. 

\begin{figure}[ht]
   \centering
  \begin{subfigure}{\linewidth}
 \includegraphics[width=0.85\linewidth]{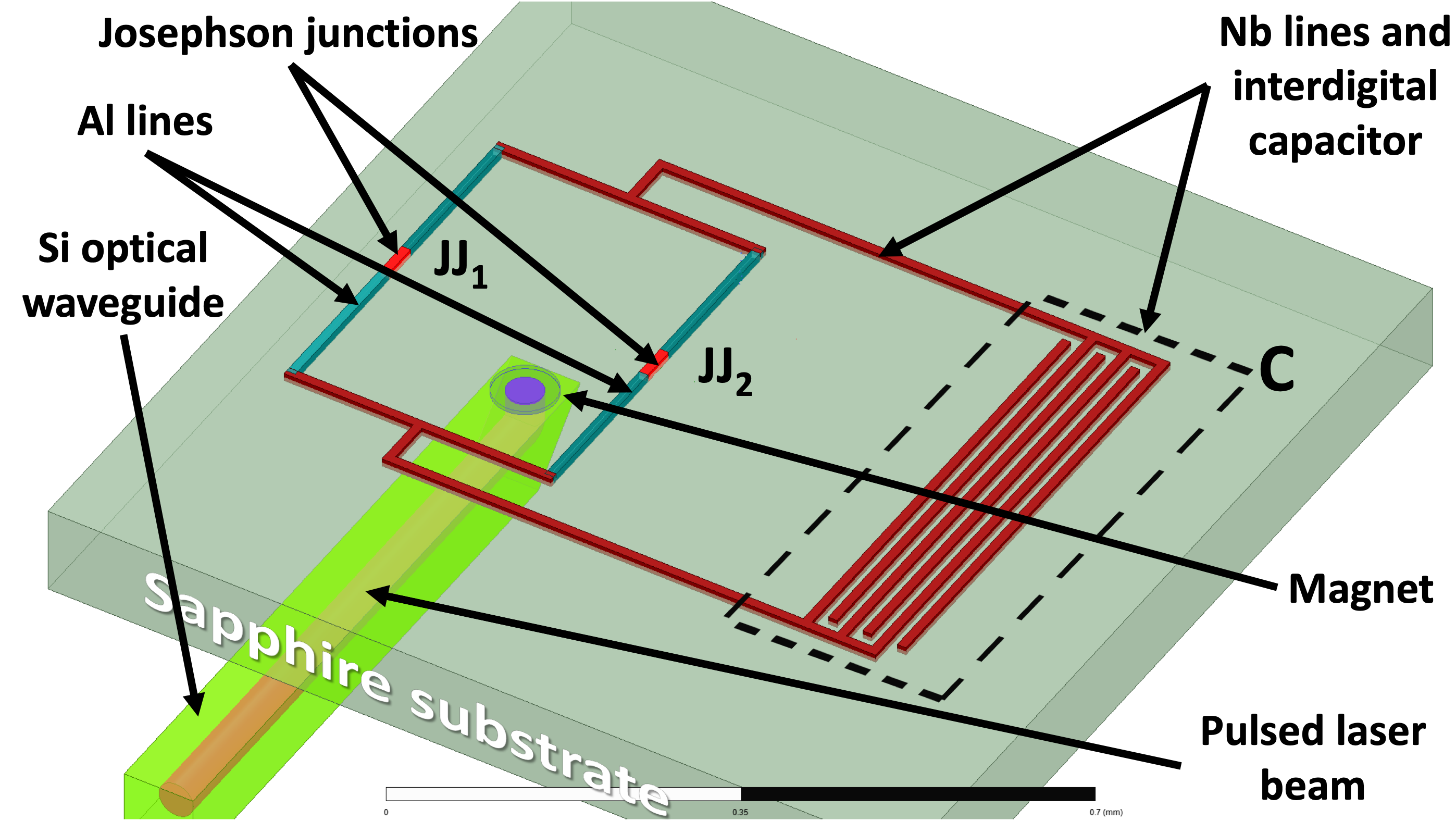}
    \caption{}
  \end{subfigure}
      \begin{subfigure}{\linewidth}
   \includegraphics[width=0.7\linewidth]{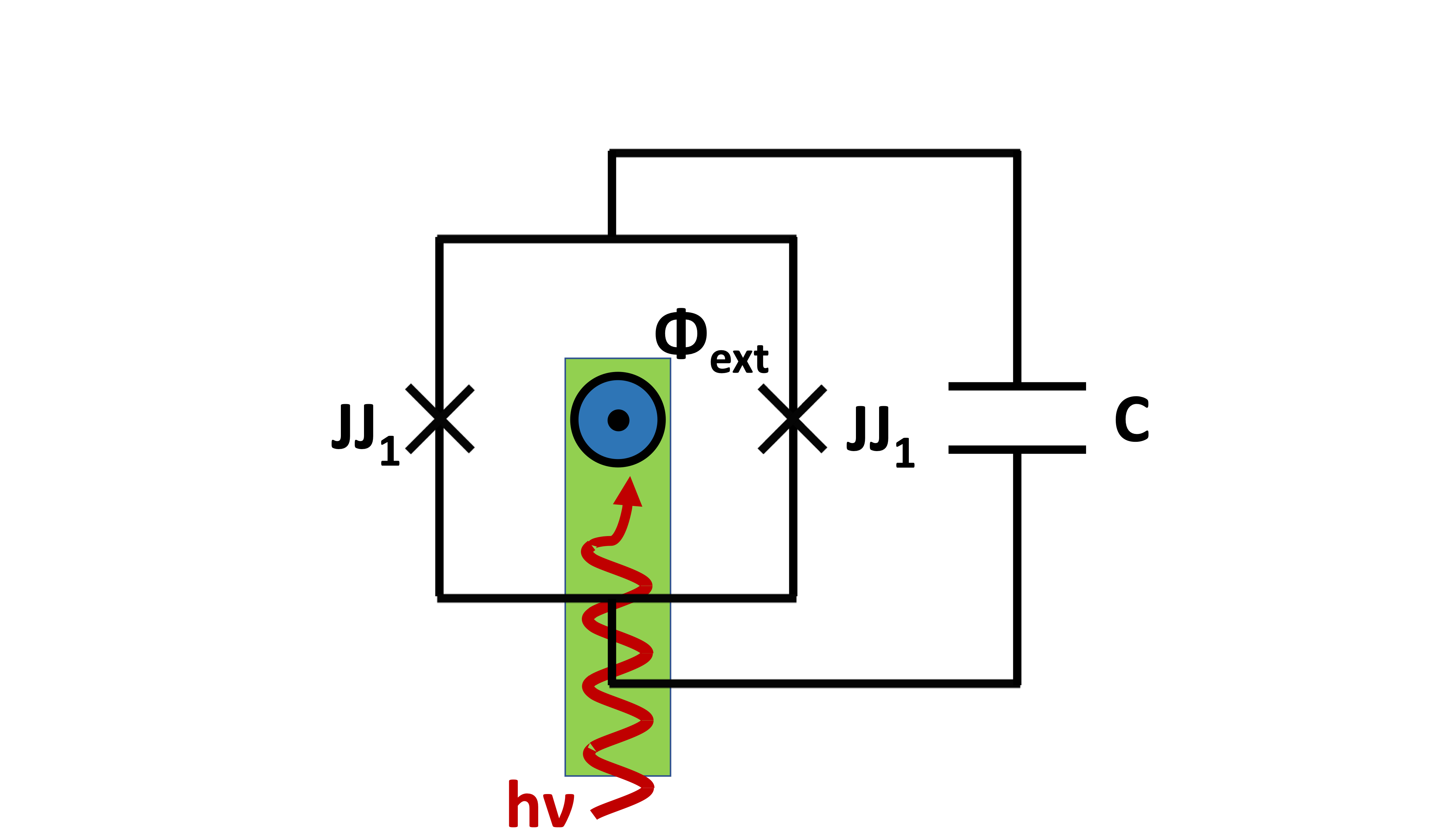}
  \caption{}
 \end{subfigure}
    \caption{(a) Schematic view of the proposed flux-tunable qubit controlled by the optically-switched magnet.a) Representation of the device and its elements (not to the scale). (b) Schematic view of the device, the external magnetic flux $\Phi_{ext}$ is generated by the optically-controlled magnet. The magnet and the optical waveguide are placed on the side of a sapphire chip that is opposite to the side with the qubits. This is done simplicity of the design and compatibility with the fabrication processes.} \label{OPT}

\end{figure}

In the envisaged architecture, see Figure \ref{OPT}, nanomagnets are controlled by means of femtosecond IR or near-IR optical pulses delivered through optical waveguides that are fabricated from low-loss dielectric materials such as Si or sapphire. Magnetization of the micro-/nanomagnets remains stable between the control optical pulses providing DC magnetic flux bias to the qubits during operation and tuning. The total losses in the system are kept at an acceptable level, as the dimensions are small and because of the ability to spatially separate the normally conducting nanomagnets away from the qubits where they still provide the necessary magnetic flux bias to qubits. Among the advantages of the described above architecture, there are compatibility with qubit nanofabrication process, scalability, and addressability of individual qubits.

\subsection{Quantum transduction}
Superconducting quantum circuits play a critical role in quantum computing and various quantum applications. However, the cryogenic operation condition restricts the capability to transfer quantum information over long distances and to build a large-scale quantum networks that connect individual nodes. One approach to break this limitation is to leverage optical signals as information carriers. These signals which operate at frequencies of hundreds of \SI{}{\tera\hertz} and are immune to thermal noise at room temperature. As such, optical photons traveling in free space or optical fibers provide promising strategies for long-distance quantum communication. The integration of such communication systems with local superconducting quantum processors requires a quantum transducer that can efficiently convert quantum signals between microwave and optical frequencies with high fidelity. 

So far, a variety of approaches have been exploited to mitigate the huge frequency gap between microwave and optical photons, by leveraging intermediate degrees of freedom such as piezo-optomechanics \cite{jiang2020efficient}, magnons \cite{hisatomi2016bidirectional}, cold atoms \cite{han2018coherent}, and rare-earth ions \cite{bartholomew2020chip}. Another promising approach for frequency conversion is by utilizing electro-optic nonlinearity which can be found in electro-optic crystals such as lithium niobate (LiNbO3 or LN) and aluminum nitride (AlN). The microwave signal applied to the crystal modulates the optical refractive index leading to a three-wave mixing process between the optical pump, the optical signal, and the microwave field. In this way a bidirectional conversion can be realized between microwave and optical signals. 

In the past years, electro-optic frequency conversion has been studied in both two-dimensional and three-dimensional architectures. In two-dimensional systems, microwave cavities are integrated with electro-optic photonic resonators on chip \cite{holzgrafe2020cavity,xu2021bidirectional}. The thin-film lithium niobate fabrication technique enabled ultrahigh qualify factor for optical resonators. In 3D platforms, bulk lithium niobate crystals are embedded into three-dimensional microwave cavities \cite{rueda2016efficient,hease2020bidirectional}, where the large mode volume and heat capacity reduce the over-heating when pump power is high. However, so far, the efficiency for quantum transduction is not high enough and most schemes operate in the high-pump regime with large noise. The reason lies in small single photon microwave-optic coupling coefficient ($g$) and low-quality factors of microwave cavities. 

To improve the quantum transduction technology, our strategy exploits Fermilab’s 3D bulk niobium cavities which have high density of the electromagnetic fields in a large RF volume with Q factor $\sim 10^{10}$. Such devices can significantly enhance the microwave-optical interactions with much lower dissipation and overheating of the devices. The large flexibility of cavity geometry can also provide new degrees of freedom to optimize the microwave-optical coupling strength. 

\begin{figure}[ht]
   \centering
  \begin{subfigure}{0.45\linewidth}
 \includegraphics[height=3 cm]{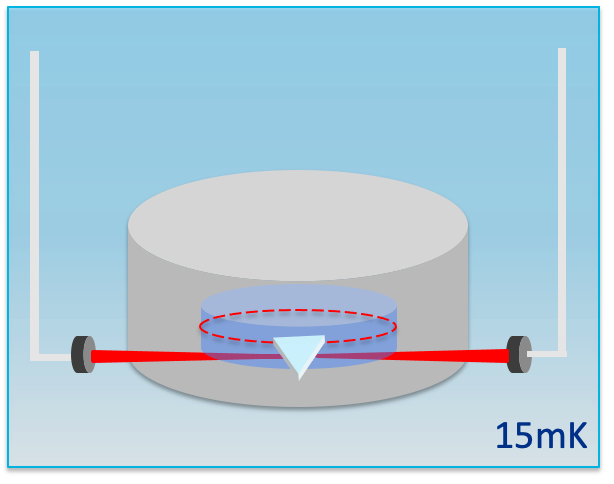}
    \caption{}
   \label{fig:cavity_block_diagram}
  \end{subfigure}
      \begin{subfigure}{0.45\linewidth}
   \includegraphics[height=3 cm]{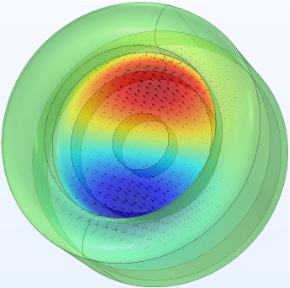}
  \caption{}
 \label{fig:dipolemode}
 \end{subfigure}
    \caption{(a) Block diagram of a microwave-optical bulk transduction device; (b) microwave design, dipole mode excited in the cavity with maximum electric field on the edge of the crystal.} 
\end{figure}

Based on the high-Q microwave cavities, we can design a new electro-optic microwave-optic frequency converter for efficient quantum transduction. The device is composed of a bulk LN crystal embedded into a high-Q bulk niobium microwave cavity, integrated with both microwave and optical couplers. The optical crystal with highly smooth surface supports whispering gallery mode where the optical wave travels along the edge of the crystal. The whole setup operates at about 15 mK temperature in a dilution refrigerator. The high microwave Q factor and large coupling strength between microwave and optical modes are expected to lead to orders of magnitude enhancement for transduction efficiency at a low pump power of tens of \SI{}{\micro\watt}. A preliminary study of this transduction technology is currently being supported through the Fermilab's Laboratory Directed Research and Development (LDRD) program. Figure \ref{fig:cavity_block_diagram} shows the microwave design of a Hybrid Quantum System (HQS). Figure \ref{fig:dipolemode} shows a preliminary microwave design, and dipole electric-field distribution.

The high-efficiency electro-optic quantum transducers lay the foundation for various applications such as high-fidelity entanglement generation and quantum sensing. For example, by performing the detection of optical photons, one can herald the generation of entangled microwave photons in two distant superconducting quantum units \cite{krastanov2021optically}. The high-quality factors of the three-dimensional microwave cavities could significantly improve the entanglement generation rate as well as the fidelity. On the other hand, one can leverage the optical homodyne detection for readout of weak microwave signals, with the measurement precision below the quantum limit \cite{nazmiev2022back}. 

\section{Room temperature hardware for quantum devices}\label{room_hardware}
The proposed 2D and 3D quantum computers consist of superconducting qubits coupled with superconducting resonators. The states of such systems are typically manipulated by feeding them finely tuned electrical radio frequency (RF) pulses. These pulses interact with the different constituent quantum elements (according to the chosen frequencies of the input signals), which in turn alter the pulses in nontrivial fashions. These new signals, which contain information about the system, can then be measured via readout pulses (see Ref. \cite{quantumengineersguide} for a comprehensive review). The control and readout pulse generation processes call for sophisticated room temperature control electronics.

The fundamental frequencies of the superconducting quantum systems typically lie in the \SI{1}{} - \SI{10}{\giga\hertz} band \cite{romanenko2020}, with the resonance frequencies of transmons being of the same order. To create such arbitrary pulses with an arbitrary waveform generator (AWG), it would need to operate at these frequencies.
If the frequencies of the generated pulses are all derived from the same reference clock (whether on an AWG of another instrument), they should, in theory, be phase aligned (although slight timing offsets between each channel’s circuitry, referred to as the skew may exist). The importance of phase-aligned signals will be discussed in the next sub-section.

Modern top-of-the-line AWGs can achieve needed bandwidths and other specifications, and some are well suited to generating the short, precise, arbitrary pulses required in quantum experiments. 
Significantly less specialised electronics will most likely be inadequate. However, such ready-made solutions are often expensive and inflexible. Indeed, given the nascent nature of the projects being carried out at SQMS - and quantum information science in general - the design specifications for the proposed 2D and 3D quantum computers will certainly change and evolve as new bottlenecks are discovered and physical hardware requirements encountered. As such, an equally nimble electronic hardware solution may be preferable to a proprietary one. The obvious choice for such a prototype electronic control system is hence one based on field programmable gate arrays (FPGAs). 
The versatility of FPGAs, coupled with the fact that these semiconductor circuits can solve any computable problem (and be programmed to implement complex logic functions and act as interconnected memory blocks, digital signal converters, entire microprocessors, etc.), means that a custom, in-house approach should be explored too. Indeed, Fermilab has itself recently developed a custom qubit control module, QICK \cite{stefanazzi2021qick}. Developing it further and integrating it into the entire hardware stack in the SQMS center is being actively considered.

Finally, the electronics of choice would mostly likely need to possess functionality to interface with a wide array of other devices for synchronization  and data transfer purposes, and as such needs to be of sufficiently high throughput and possess sufficient interfaces. However, it should be noted that phase alignment between different instruments, if required, is usually complex.

\subsubsection*{Specification requirements for 3D QPUs}

The 3D SRF/transmon quantum computer being developed by the SQMS Center is incredibly novel and offers a number of advantages over more traditional, solely planar qubit architectures, such as all-to-all connectivity and record-high coherence times.

The parameters of said pulses, namely their amplitudes, frequencies, phases, and overall shape, require precise control. Indeed, the area under a pulse as well as its phase relative to that of all other signals determines the quantum operation that is applied to the system \cite{quantumengineersguide}. These arbitrary pulses are generated using mixing as described earlier. Now, while such mixing is standard in commonplace electronics, the relatively high resolutions, low latencies, and low error tolerances required to control SQMS' 3D quantum computer calls for specialist electronics. First, one which avoids complicated external analog mixer calibration and instead opts for high frequency direct digital synthesis, along with digital IQ mixers, would ameliorate potential errors. Furthermore, such a control system would be more amenable to hardware acceleration in attempts to reduce overall system latency. This is important because the decision-making logic often needs to occur in real time. For example, so-called active reset protocols involve measuring a qubit (or cavity in a 3D architecture) and applying a ‘reset’ pulse if the qubit is found to be in an excited state \cite{PhysRevLett.110.120501}. This needs to occur before other qubits or cavity states evolve and potentially decohere; empirically, the total time for this entire feedback loop needs to be on the order of \SI{100}{\nano\second}. To best leverage SQMS’s long coherence cavity times, the hardware itself cannot be the bottleneck: if such feedback instructions were carried out chiefly in software (on a desktop computer interfacing with the qubit controller for example), latencies of up to $\sim$ \SI{1}{\milli\second} may be introduced; empirically, we have observed latencies of 3 to 4 orders of magnitude less on hardware accelerated, FPGA-based systems. Similarly quick \textit{measure-decide-generate-pulse} protocols feature in error corrections schemes too and will feature prominently in the current NISQ era.

In the 3D architecture, as in most superconducting quantum computers, it is fundamental that the various control and readout signals are phase-locked over large bandwidths to properly keep track of the phase relationships between the quantum subsystems. To see why, consider the case of a single transmon qubit. The relative phase between the computational basis state vectors of the transmon is constantly evolving, at its respective frequency, as the state rotates around the Bloch sphere of the qubit in the laboratory reference frame \cite{quantumengineersguide}. When a pulse is applied to the qubit, the phase of the pulse and its relationship with the phase of the qubit determines the quantum gate that that pulse corresponds with. For example, if the qubit is in the $\ket{0} + \ket{1}$ state the moment a $\pi$ pulse is applied, the qubit is moved to the $\ket{0} - \ket{1}$ state. However, as the qubit state is simultaneously rotating about the equator of the Bloch sphere, if the $\pi$ pulse is inadvertently applied a quarter of a period too late (when the qubit is in the $\ket{0} + i\ket{1}$ state), the end state of the qubit would be $\ket{0} -i \ket{1}$. Now, phase-locking would not be important in the case of a single transmon, as one need simply time the pulse sequence correctly to implement the correct single qubit gate. However, the phases between the control signals for multiple qubits need to be locked with respect to one another so that the phases can be precisely accounted for to properly affect various two qubit gates. For example, if a series of pulses without phase-locking (and hence probable mismatch), with which we wish to implement a CNOT gate, are implemented, the mismatch between the signals would affect the axis about which one (or both) qubits are rotated on their respective Bloch spheres, having a deleterious effect on the process fidelity. The same concerns carry over to a 3D system comprised of transmons and cavities. So, electronics which phase-lock the signals with respect to a reference signal are hence needed. Furthermore, this phase-locking needs to occur over a large bandwidth, given the high frequencies at which the constituent elements are required to operate.

Another challenge that needs to be addressed in order to control a very large Hilbert space is the level of selectivity, which allows for the preparation of any arbitrary superposition of Fock states and ultimately quantum information processing. Controls of a large Hilbert space requires the ability of controlling multiple microwave signals over a wide spectrum and with high resolution in frequency and amplitude. An additional challenge relies in the synthesis of OQC pulses, that may requires fast transitions over a large bandwidth. Error correction schemes based on feedback loops are well known and already employed, however they mostly are implemented offline. Most commercially available platforms cannot reach this level of performances, further development in this field along with the implementation of active loops with low latency would further facilitate the controllability of 2D and 3D quantum devices. The implementation of fast Digital to Analogue (DAC) converters flexible digital signal processing blocks to allow live processing, data acquisition and analysis, will also represent a breakthrough to enhance performances and improve readout fidelity.

\section{Quantum error protection and correction}\label{error_correction}
To execute any of the advanced quantum algorithms the quantum processors need to be reliable throughout the computation, i.e. during gate operation and information storage. Due to the interaction with the environment (albeit passive or during control operations), errors are however inevitable, and a good understanding of their origin is required to make the execution of quantum algorithms sufficiently robust. In general, there are three separate strategies to reduce the rate at which quantum information is adversely affected by errors i.e 1) decrease the coupling to the noisy environment by filtering and shielding, 2) adopt a circuit design with eigenstates immune to noise and 3) actively measure the presence of errors and correct accordingly. These strategies are complementary with each other, and thus can be implemented together to enhance the performance of the quantum processor. In general, the strategy to reduce the adverse effects of any spurious coupling to the environment is referred to as quantum error correction. In this section both the passive (hardware-level) protection and active (feedback based) error correction will be discussed. 

\subsection{Error-protected quantum devices} 
Quantum information processing relies on the encoding of information into the superposition of a systems eigenstates e.g. $\ket{\psi} = \cos(\theta/2)\ket{0} + e^{i\phi}\sin(\theta/2)\ket{1}$. In contrast to classical information processing, such an encoding is determined by complex instead of real coefficients. Consequently, quantum hardware deals with two types of errors i.e. in addition to bit-flips (related to $\theta$) caused by relaxation, the information can be corrupted by dephasing due to phase-flips (related to $\phi$). The rate of both errors is determined by specific properties of the eigenstates set by the device circuit. The dephasing rate, for example, depends on the energy dispersion of the eigenstates with respect any of the device parameters i.e. 

\begin{equation}
1/T^{\lambda}_{\phi} \propto \bigg| \frac{\partial E_{01}}{\partial \lambda} \bigg|^2 S_{\lambda} ( \omega \rightarrow 0)
\end{equation}

where $S_{\lambda}(\omega)$ is the power spectral density of the noise related to device parameter $\lambda$. This relation is at the root of the successful transmon device as its energy dispersion is exponentially suppressed with the $E_J/E_C$ ratio. The transmon relaxation rate, one the other hand, is not particularly minimized. In general the relaxation rate is determined by Fermi's Golden rule for a specific noisy interaction operator $\hat{\mathcal{O}}$ i.e. 

\begin{equation}
    1/T^{\lambda}_{1} \propto |\bra{0} \hat{\mathcal{O}} \ket{1}|^2 S_{\lambda}(\omega = E_{01}/\hbar)
\end{equation}

where $E_{01}$ is the energy difference between the two eigenstates. Fluxonium qubits, for example, typically have higher coherence times compared to other superconducting circuits \cite{Manucharyan_1ms_fluxonium} due to the smaller charge transition matrix element. Additionally, they have a much larger, and positive anharmonicity which potential is beneficial for its operation in both a 2D based as 3D based quantum architecture.  Despite a slightly more complicated circuit geometry including a necessary flux bias control, the two major advantages mentioned still drive both the academia and industry to increasingly invest in this type of device \cite{nguyen2022scalable}. Recently the fidelities of two-qubit gates on the fluxonium has reached the same level of those of the transmons device, which makes it a strong candidate for the 2D based architecture.

The SQMS project pursues further decreasing the fluxonium's dephasing and relaxation rate. One indispensable step toward this goal is to more carefully characterizing the noise channels. On the dielectric loss and 1/$f$ flux noise, it is still unclear what are the main parameters impacting the noise amplitudes. Possible factors include the circuit size, geometry, and materials. Finally, the presence of quasi-particles and the instability of critical current should also add to the decoherence of the fluxonium qubit are part of further research. 

The transmon and fluxonium circuits are both examples of devices with (partial) protection of either depolarization or pure dephasing noise. However, the coherence time $T_2$ of a device is determined from both depolarization $T_1$ and pure dephasing time $T_\phi$ as $1/T_2 = 1/2 T_1 + 1/T_\phi$. Clearly, a qubit with simultaneous protection against both types of noise is required in order to have long coherence time. One candidate is the so-called $\cos 2\varphi$ qubit~\cite{Doucot2002}. The key element in this ideal circuit is a physical system which only allows tunneling of \textit{pairs} of Cooper pairs. 
By encoding logical states in subspaces with even and odd parity of the Cooper pairs, the states exhibit disjoint support in the charge basis, and thus protect the qubit from depolarization noise such as dielectric loss. 
In order to reduce charge noise, which is the dominant dephasing noise, we simply take large $E_\text{J}/E_\text{C}$ ratio, similar to the strategy used in the transmon qubit. 
It should be emphasized that the $\cos{2\varphi}$ qubit introduced here is an \textit{ideal} circuit, in the sense that there is no physical system with a natural potential with $\pi$-periodicity. The $0-\pi$ qubit, first introduced in Refs.~\cite{Doucot2002,ioffee2002,Kitaev2006,Brooks2013,Dempster2014,Groszkowski2017a,DiPaolo2018a} and recently realized in Ref.~\cite{PRXQuantum.2.010339}, can be viewed as a physical implementation of the $\cos{2\varphi}$ qubit. With large impedance associated with superinductors in the $0-\pi$ qubit, the circuit dynamics is dominated by co-tunneling of Cooper pairs, and thus an effective $\cos{2\varphi}$ qubit. The first experimental realization exhibits an outstanding $T_1= $ \SI{1.6}{\milli\second}, and a reasonable $T_2=$ \SI{25}{\micro\second}, demonstrating the promising future of protected qubits.

A dual circuit to the $\cos{2\varphi}$ qubit is the Bifluxon qubit~\cite{Kalashnikov2019}. Instead of tunneling of pairs of Cooper pairs, it exploits tunneling of pairs of fluxons into the circuit loop, which conserves the fluxon parity, and thus enables disjoint support. With large $E_\text{C}/E_\text{L}$ ratio, the qubit can be also protected against flux dephasing noise. 

The inevitable coupling between the quantum system to its environment in general leads to the loss of information and degrade the performance of QPUs. However, there are cases where the noise is helpful, especially when a drive is applied to the system. The strategy to engineer the Hamiltonian and the dissipation of the system by driving is referred as driven-dissipative. In the context of superconducting circuits, this often involves a nonlinear element (e.g., Josephson junctions) which enables three- or four-wave mixing process. One great example here is the stabilized cat qubit. The cat qubit will eventually decay to vacuum if there is no photon injection. Thus, it is crucial to stabilize the cat states. One approach (dissipative cat) is taken in Refs.~\cite{Mirrahimi2014,Leghtas2015,Touzard2018,Lescanne2020}, where they engineer both a two-photon drive and a two-photon decay channel. This results in a decoherence-free subspace spanned by the cat states. Another direction (Kerr cat) is explored in Refs.~\cite{Puri2017,Puri2018,Grimm2020a}, where only the Hamiltonian is engineered by introducing a two-photon drive. Along with the Kerr non-linearity, the cat states can also be stabilized. Moreover, this stabilization protocol also protects the qubit from bit-flip error, as long as the two cat states are further apart in the resonator phase space, thus resulting in exponentially long lifetime. 

This class of novel qubits circuits with built-in protection against relaxation and/or dephasing clearly has a great potential for 2D processors and even as an ancilla qubit in cavity based QPUs. While the relaxation time of SRF cavities are typically still much higher than those of protected qubits, it will be reduced by the ancilla qubit via the reverse Purcell effect. This puts a limitation to the maximum coupling between the cavity and the qubit which, as shown above, translates to a minimum pulse length for SNAP gates. Simultaneously, any increase in coherence of the ancilla qubit will directly result in a larger Hilbert space available for qudit encoding. The requirement of a (static) magnetic field does however introduce serious technical complications since any magnetic field can easily reduce the relaxation time of niobium cavities. This technical challenge can be resolved by cleverly engineering of the qubit circuit, the magnet source (e.g. with broadband inductive properties) and the cavity design and material. Considering the state-of-the-art performance of the transmon qubit and the potential of protected qubit, it will become relevant to focus future research on the development and integration of protected qubits to resolve these issues. 

\subsection{Feedback error correction}
The hardware capabilities described in this paper will eventually be limited by the noisy environment. 
The HEP and CM simulation algorithms are also going to be affected by this noise. 
Besides the hardware efforts for the error correction, a theory-based error correction is needed. 
The bosonic error correction can be done via enlargement of the Hilbert space, then extracting the information from the redundant information \cite{shor1995scheme, gottesman2010introduction, cai2021bosonic}. 
The error correction protocols can change depending if the information is coded in cat (coherent) states or Fock states. For the error correction on Fock states, the GKP (Gottesman, Kitaev, Preskill) code can be utilized \cite{gottesman2001encoding}. The GKP codes take advantage of displacement gates by using them as stabilizer in the position-momentum phase space, if the errors are $|\delta q|<\sqrt{\pi}/2$ and $|\delta p|<\sqrt{\pi}/2$ \cite{cai2021bosonic}. 
The GKP code can be utilized in SNAP and displacement based HEP simulation algorithms. 
The accuracy of the GKP code when the qudit dimension gets large is another challenge that needs to be addressed. 
For large error rates and large qudit dimensions, surface codes can be utilized to correct the bosonic errors \cite{fowler2012surface}.

\section{HEP cloud}\label{HEPcloud}
As SQMS hardware develops and matures, there will be an increasing need to standardize interfaces and procedures to conduct experiments and studies using Fermilab quantum devices. Through the HEPCloud1system\footnote{https://hepcloud.fnal.gov}, Fermilab provides to the user community common methods to access computational resources available on site and offsite. HEPCloud has features for scheduling job, reserving resources  and managing users and time allocations. Integrating quantum computing into HEPCloud began in 2021. We have successfully demonstrated HEPCloud scheduling of batch jobs originating at Fermilab and returning results back to the user space at Fermilab. We are currently working to define and develop the interfaces necessary to access QPUs. This HEPCloud-quantum computing effort also serves as the groundwork for integrating the experimental devices that will be available at Fermilab through SQMS over the next few years. We anticipate adding features for handling data collection, data cataloging, and access. HEPCloud can also allow for expansion into High Performance Computing facilities at NERSC\footnote{National Energy Research Scientific Computing Center} and ALCF\footnote{Argonne Leadership Computing Facility} when there is a need for high computational load. The same interfaces used for device access can also be used for quantum simulations and pulse design with resource access controls and accounting managed centrally for simplified user access. 

\section{Summary and Outlook}\label{Conclusions}

In this whitepaper we reviewed most recent developments in quantum computing based on superconducting technology, highlighting the synergies between the QIS and HEP programs, as well as future research directions. 
Some of the R\&D challenges are outlined as follows.

\begin{itemize}
    \item \textit{HEP Simulations in 2D and 3D devices} -- With fermionic modes of the transmon and bosonic modes of the electric field in a cavity, we provide alternative and potentially powerful techniques to the qubit based HEP simulation devices. With the engineering of the alternative gates and new discretization techniques, various field theories can be simulated on the SRF devices that are developed at SQMS. Gate engineering, gate and readout fidelities, new encoding schemes are the major challenges for HEP simulations on SRF platforms. 
    
    \item \textit{Quantum hardware} -- 
    The realization of superior quantum devices still requires to overcome several challenges. For the qutrit architecture these are mostly related to materials. The qudit architecture however is less mature and more flexibility which justifies further research into optimal information encoding, qudit-qudit connectivity and operation. 
    
    \item \textit{Noise-protected qubits} -- 
    We propose to go beyond the current transmon qubit and explore the possibility of 2D qubits with intrinsic noise protection, such as the fluxonium and $\cos2\varphi$ qubit. 
    Served as a processing unit or an ancilla, the protected qubit can improve the performance in both 2D and 3D QPU. 
    
    \item \textit{Room temperature electronics and instrumentation} -- Development and implementation of superior room temperature electronics, with low latency, reduced phase noise and enhanced resolution. This would also further facilitate the controllability of 2D and 3D quantum devices over a large Hilbert space.
    
    \item \textit{Quantum networks} -- Advances in the realization of distributed quantum networks is probably the next major challenge to be addressed. Whether it is realized through transduction, hybrid or microwave lines, any progress in this field is a step toward the scaling of current quantum processing units and sensors.
    
    \item \textit{Quantum state transfer (QST)} -- Distributed quantum information processing among spatially separated nodes is a key ingredient for scalable quantum computing.  We plan to achieve this by demonstrating on-demand high-fidelity quantum state transfer (QST) based on SQMS-specific infrastructure.  Major components, including node configuration, node-transmission channel coupling, coupling modulation mechanism, pulse shaping protocols, etc. will undergo design, prototyping, testing and integration.  Staged approach, with transmission channel going from milli-Kelvin to a few Kelvins, is envisioned aiming for breakthrough possibilities.

\end{itemize}

In general, we find there is ample justification to expect breakthrough advances in HEP algorithms, interconnectivity and quantum hardware performance, especially for the qudit based architectures.

\section{Acknowledgements}

This material is based upon work supported by the U.S. Department of Energy, Office of Science, National Quantum Information Science Research Centers, Superconducting Quantum Materials and Systems Center (SQMS) under the contract No. DE-AC02-07CH11359. Fermilab is operated by the Fermi Research Alliance, LLC under contract No. DE-AC02-07CH11359 with the United States Department of Energy. 
Sohaib M. Alam and Davide Venturelli are supported by the NASA Academic Mission Services, Contract No. NNA16BD14C. The authors would like to thank Srivatsan Chakram for the valuable discussion and feedback on topics related to this whitepaper.

\clearpage

\end{document}